\documentclass[11pt,a4paper]{article}
\usepackage{amsmath}
\usepackage{amsfonts}
\usepackage{diagbox}
\usepackage{array}
\usepackage{bm}
\usepackage{tikz}
\usetikzlibrary{fit}					
\usetikzlibrary{backgrounds}	%

\usepackage{mathrsfs}
\usepackage{cite}

\newcommand*\xbar[1]{%
  \hbox{%
    \vbox{%
      \hrule height 0.5pt 
      \kern0.5ex
      \hbox{%
        \kern-0.1em
        \ensuremath{#1}%
        \kern-0.1em
      }%
    }%
  }%
}

\begin{document}

\title{The gauge coupling unification in the flipped $E_8$ GUT}

\author{
K.V.Stepanyantz${}^{abc}$ $\vphantom{\Big(}$
\medskip\\
{\small{\em Moscow State University, Faculty of Physics,}}\\
${}^a${\small{\em Department of Theoretical Physics,}}\\
${}^b${\small{\em Department of Quantum Theory and High Energy Physics,}}\\
{\small{\em 119991, Moscow, Russia}}\\
\vphantom{1}\vspace*{-2mm}\\
$^c$ {\small{\em Bogoliubov Laboratory of Theoretical Physics, JINR,}}\\
{\small{\em 141980 Dubna, Moscow region, Russia.}}\\
\vphantom{1}\vspace*{-2mm}}

\maketitle

\begin{abstract}
The gauge coupling unification is investigated at the classical level under the assumptions that the gauge symmetry breaking chain is
$E_8\to E_7\times U_1 \to E_6\times U_1 \to SO_{10}\times U_1 \to SU_5 \times U_1 \to SU_3 \times SU_2 \times U_1$ and only components of the representations 248 of $E_8$ can acquire vacuum expectation values. We demonstrate that there are several options for the relations between the gauge couplings of the resulting theory, but the only symmetry breaking pattern corresponds to $\alpha_3=\alpha_2$ and $\sin^2\theta_W=3/8$. Moreover, only for this option the particle content of the resulting theory includes all MSSM superfields. It is also noted that this symmetry breaking pattern corresponds to the case when all representation which acquire vacuum expectation values have the minimal absolute values of the relevant $U_1$ charges.
\end{abstract}

\section{Introduction}
\hspace*{\parindent}

Quantum numbers of various fields in the Standard Model are not random. For example, they satisfy some highly nontrivial constraints following from the anomaly cancellation condition, see, e.g., \cite{Minahan:1989vd,Bilal:2008qx}. Moreover, many years ago it became clear \cite{Georgi:1974sy} that they imply the existence of a wider gauge symmetry, so that the Standard model presumably appears as a low energy remnant of a more symmetric Grand Unified theory (GUT) \cite{Mohapatra:1986uf}. The simplest GUT \cite{Georgi:1974sy} has the gauge group $SU_5$ and predicts unification of the gauge couplings $\alpha_2 = \alpha_3$ and the value of the Weinberg angle $\sin^2\theta_W=3/8$. These predictions should be valid at very high energies of the order of the $SU_5$ symmetry breaking scale (about $10^{16}$ GeV). If the coupling constant for the hypercharge $U_1$ group is denoted by $e_1$, then it is convenient to introduce the coupling

\begin{equation}\label{Alpha1_Definition}
\alpha_1\equiv \frac{5}{3}\cdot \frac{e_1^2}{4\pi},
\end{equation}

\noindent
because in this case the gauge coupling unification condition can be written in the simple form $\alpha_1=\alpha_2=\alpha_3$. Although the running of couplings in the Standard model does not match this prediction, magically, this condition agrees with the running of couplings in its simplest supersymmetric extension called the Minimal Supersymmetric Standard Model (MSSM) \cite{Ellis:1990wk,Amaldi:1991cn,Langacker:1991an}. Of course, this fact is hardly accidental.

However, the simplest supersymmetric GUT based on the gauge group $SU_5$ has some essential problems. For instance, the doublet-triplet splitting requires fine tuning \cite{Dimopoulos:1981zb,Sakai:1981gr}, and the predicted proton lifetime is less than the current experimental restrictions \cite{Workman:2022ynf}. These problems are successfully solved in supersymmetric GUTs with the extra space dimensions compactified on orbifolds \cite{Kawamura:2000ev,Altarelli:2001qj,Hall:2001pg,Kobakhidze:2001yk}, see also \cite{Raby:2017ucc}. However in this paper we will consider usual supersymmetric GUTs in four dimensions.

In the simplest $SU_5$ theory the left fermions of a single generation\footnote{Standardly (see, e.g., \cite{Cheng:1984vwu}), they contain the charge conjugated right fermions of the Standard model and the charge conjugated right neutrinos.} belong to the reducible representation $1+\xbar{5}+10$ of $SU_5$. Equivalently, it is possible to deal with the right fermions in the representation $1+5+\,\xbar{10}\,$. In particular, in the supersymmetric case chiral superfields of a single generation can be written the form

\begin{equation}\label{Single_Generation}
\,\xbar{10}\,^{ij} \sim \left(
\begin{array}{ccccc}
0 & U_3 & -U_2 & \widetilde U^1 & \widetilde D^1\\
-U_3 & 0 & U_1 & \widetilde U^2 & \widetilde D^2\\
U_2 & -U_1 & 0 & \widetilde U^3 & \widetilde D^3\\
-\widetilde U^1 & -\widetilde U^2 & - \widetilde U^3 & 0 & E\\
-\widetilde D^1 & -\widetilde D^2 & -\widetilde D^3 & -E & 0
\end{array}
\right);\qquad 5_i \sim \left(\begin{array}{c} D_1\\ D_2\\ D_3\\ \widetilde E\\ -\widetilde N
\end{array}\right);\qquad 1\sim N,
\end{equation}

\noindent
where $N$, $E$, $U$, and $D$ include the right neutrino, charged lepton, up and down quarks, respectively, and the charge conjugated left leptons and quarks are components of the superfields

\begin{equation}
\left(
\begin{array}{c}
\widetilde N\\
\widetilde E
\end{array}
\right);\qquad
\left(
\begin{array}{c}
\widetilde U\\
\widetilde D
\end{array}
\right),
\end{equation}

\noindent
respectively. By a vacuum expectation value of the Higgs field in the adjoint representation $24$ the $SU_5$ symmetry can be broken down to the subgroup $SU_3\times SU_2\times U_1$ with the elements

\begin{equation}
\left(
\begin{array}{cc}
\omega_3 e^{-i\beta_Y/3} & 0 \\
0 & \omega_2^* e^{i\beta_Y/2}
\end{array}
\right) \in SU_3\times SU_2 \times U_1 \subset SU_5.
\end{equation}

\noindent
Then, from the $SU_5$ tensor transformations one obtains that with respect to the subgroup $SU_3\times SU_2\times U_1$ the chiral superfields in Eq. (\ref{Single_Generation}) have the same quantum numbers as the MSSM superfields.

Wonderfully, it is the representation $1+5+\,\xbar{10}\,$ that appears in the decomposition of the $SO_{10}$ spinor irreducible representation with respect to the subgroup $SU_5\times U_1$ \cite{Slansky:1981yr},

\begin{equation}\label{16_Decomposition}
\xbar{16}\,\Big|_{SO_{10}}=1(5)+ 5(-3)+\,\xbar{10}\,(1)\Big|_{SU_5\times U_1},
\end{equation}

\noindent
where the $U_1$ charges are presented in the brackets. Therefore, the theory based on the gauge group $SO_{10}$ \cite{Fritzsch:1974nn,Georgi:1974my} is more attractive because all particles of a single generation belong to a single irreducible representation $\,\xbar{16}\,$ (or $16$). Theories based on the group $SO_{10}$ have attracted considerable attention, see \cite{Joshipura:2009tg,Patel:2010hr,Joshipura:2011nn,DeRomeri:2011ie,Babu:2011tw,Babu:2012vc,Awasthi:2013ff,Altarelli:2013aqa,Mambrini:2016dca,Babu:2016bmy,Bjorkeroth:2017ybg,Deppisch:2017xhv,Chakrabortty:2017mgi,Antusch:2017tud,
Mohapatra:2018biy,Babu:2018tfi,Ellis:2018khn,Boucenna:2018wjc,Babu:2018qca,Ohlsson:2019sja,Haba:2019wwt,Chakrabortty:2019fov,Chakraborty:2019uxk,Hamada:2020isl,Ohlsson:2020rjc,Chakrabortty:2020otp,King:2021gmj,Ding:2021eva,
Mummidi:2021anm,Cho:2021yue,Patel:2022wya,Held:2022hnw,Sahu:2022rwq,Maji:2022jzu,Lazarides:2022ezc,Haba:2022myj,Patel:2022nek,Patel:2022xxu} for some recent references.

Note that the $SO_{10}$ symmetry can be broken in different ways, and the most conspicuous symmetry breaking pattern $SO_{10} \to SU_5 \to SU_3\times SU_2 \times U_1$ is not the best one. It is much better to break the symmetry via the flipped $SU_5$ model \cite{Barr:1981qv,Antoniadis:1987dx,Campbell:1987eb,Ellis:1988tx}, so that $SO_{10}\to SU_{5}\times U_1 \to SU_3\times SU_2\times U_1$.

In this model the gauge group is $SU_5\times U_1$, and the chiral superfields of a single generation are accomodated in the representations of the group $SU_5$ in a different way in comparison with Eq. (\ref{Single_Generation}),

\begin{equation}
\,\xbar{10}\,{}^{ij} \sim \left(
\begin{array}{ccccc}
0 & D_3 & -D_2 & \widetilde U^1 & \widetilde D^1\\
-D_3 & 0 & D_1 & \widetilde U^2 & \widetilde D^2\\
D_2 & -D_1 & 0 & \widetilde U^3 & \widetilde D^3\\
-\widetilde U^1 & -\widetilde U^2 & - \widetilde U^3 & 0 & N\\
-\widetilde D^1 & -\widetilde D^2 & -\widetilde D^3 & -N & 0
\end{array}
\right);\qquad 5_i \sim \left(\begin{array}{c} U_1\\ U_2\\ U_3\\ \widetilde E\\ -\widetilde N
\end{array}\right);\qquad 1\sim E.
\end{equation}

\noindent
We see that in comparison with Eq. (\ref{Single_Generation}) the superfields $U$ and $D$, $N$ and $E$ are swapped, due to which this theory is called flipped. However, when the symmetry is broken down to $SU_3\times SU_2\times U_1$ the unbroken $U_1$ subgroup is now a superposition of the $SU_5$ transformations

\begin{equation}
\omega_5 = \exp\Big\{\,\frac{i\beta_Y}{30}\left(
\begin{array}{cc}
2\cdot 1_3 & 0 \\
0 & -3\cdot 1_2
\end{array}
\right)\Big\}
\end{equation}

\noindent
and the transformations of the $U_1$ factor in the gauge group $SU_5\times U_1$ with $\omega_1 = \exp(-iq\beta_Y/5)$, where the $U_1$ charge $q$ is normalized by Eq. (\ref{16_Decomposition}). It is easy to see that in this case one obtains correct quantum numbers for all MSSM superfields. The breaking of the $SU_5\times U_1$ symmetry can be realized if vacuum expectation values are acquired by the Higgs fields in the representations $10 (-1)$ and $\overline{10} (1)$.

The flipped $SU_5$ model is not excluded by the present limits on the proton lifetime \cite{Ellis:2020qad} (see also \cite{Mehmood:2020irm,Haba:2021rzs,Ellis:2021vpp}), allows to avoid doublet-triplet splitting problem \cite{Antoniadis:1987dx,Masiero:1982fe,Grinstein:1982um,Hisano:1994fn}, does not involve irreducible representations of very large dimensions, and can be derived from string arguments \cite{Antoniadis:1987tv,Antoniadis:1989zy}. The recent investigations of this model can be found in \cite{Huang:2006nu,Chung:2010bn,Kuflik:2010dg,Ellis:2011es,Ellis:2018moe,Ellis:2019jha,Hamaguchi:2020tet,Korneev:2021zdz,Charalampous:2021gmf,Antoniadis:2021rfm,Basiouris:2022wei,
Du:2022lij}.

Some other groups are also used for constructing of various GUTs. For instance, theories based on the gauge group $E_6$ \cite{Gursey:1975ki} (see \cite{King:2020ldn} for a recent review) deserved especial attention, because they appear after the superstring compactifications on the Calabi--Yau manifolds \cite{Candelas:1985en,Witten:1985xc}, see also \cite{Green:1987mn}. Different patterns which can be used for breaking the $E_6$ symmetry are listed in \cite{Slansky:1981yr}. The various phenomenological aspects of theories based on the group $E_6$ have been discussed in \cite{Chakrabortty:2017mgi,Chakrabortty:2019fov,Chakrabortty:2020otp,Caravaglios:2005gf,Stech:2008wd,King:2008qb,Athron:2009ue,Athron:2009bs,Das:2010cp,Hall:2010ix,Wang:2011eh,Nevzorov:2013tta,Athron:2014pua,Nevzorov:2015sha,Athron:2015vxg,Ko:2015fxa,Athron:2016gor,Nevzorov:2017rtf,
Nevzorov:2018leq,Dutta:2018qei,Nevzorov:2020uuh,Nevzorov:2022zjo,Nevzorov:2022zns,Nevzorov:2023scg}.

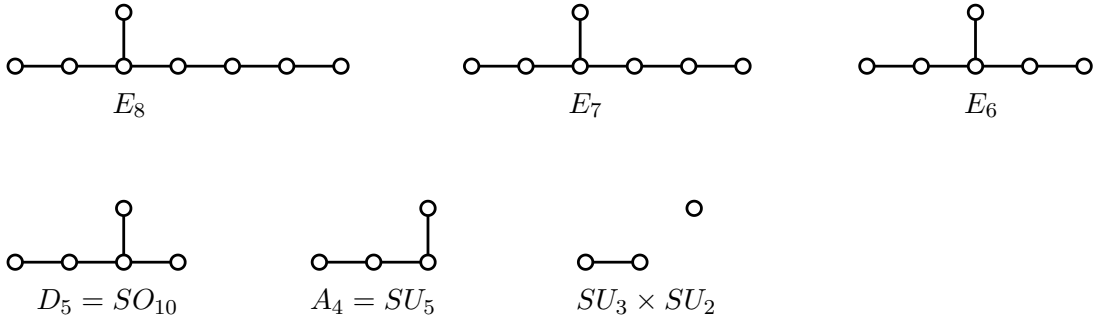
\begin{figure}[h]
\begin{picture}(0,5.3)
\put(1,3.1){\includegraphics[scale=0.25]{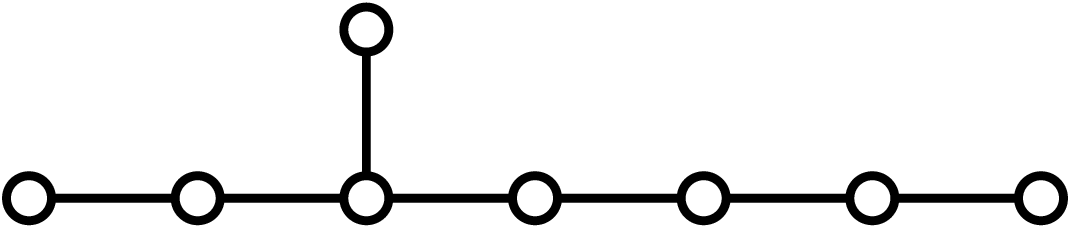}}
\put(7,3.1){\includegraphics[scale=0.25]{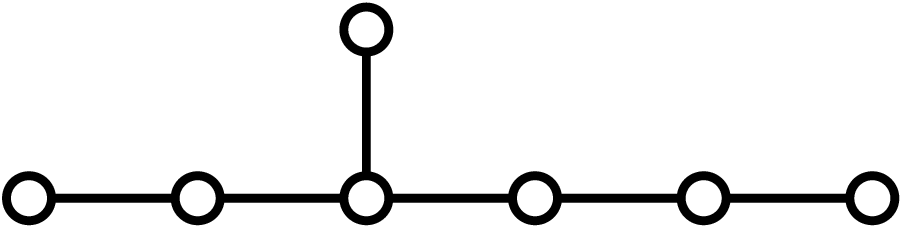}}
\put(12.2,3.1){\includegraphics[scale=0.25]{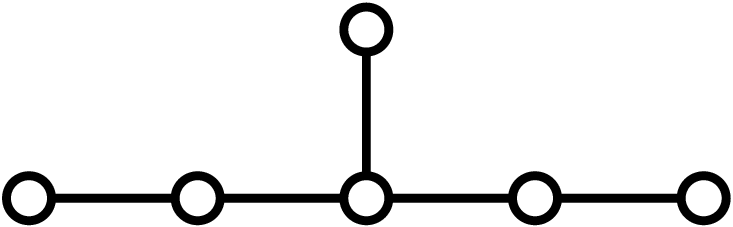}}
\put(1,0.5){\includegraphics[scale=0.25]{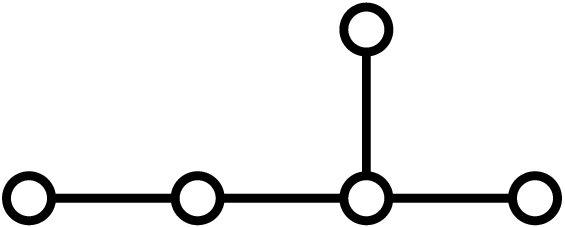}} \put(1.4,0){$D_5 = SO_{10}$}
\put(5,0.5){\includegraphics[scale=0.25]{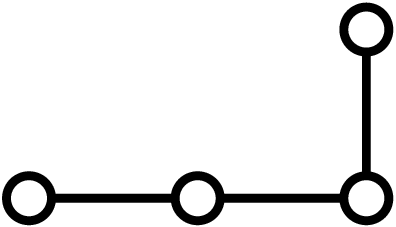}} \put(5,0){$A_4 = SU_5$}
\put(8.5,0.5){\includegraphics[scale=0.25]{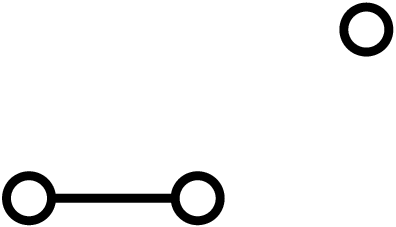}} \put(8.5,0){$SU_3\times SU_2$}
\put(2.4,2.6){$E_8$}
\put(8.4,2.6){$E_7$}
\put(13.6,2.6){$E_6$}
\end{picture}
\bigskip
\caption{Dynkin diagrams for the exceptional groups of the $E$-series. Continuing the sequence one obtains the Dynkin diagrams for $D_5=SO_{10}$, $A_4= SU_5$, and $SU_3\times SU_2$.}
\label{Figure_Dynkin}
\end{figure}

Note that the Dynkin diagrams for the groups $E_6$, $SO_{10}$, $SU_5$, and even $SU_3\times SU_2$ are very similar. Actually, all of them can be viewed as a continuation of the $E$-series, which is started from the group $E_8$. This is illustrated in Fig. \ref{Figure_Dynkin}. The similarity between these Dynkin diagrams hints that the larger groups of the $E$-series can be used for constructing GUTs. The main obstacle for this is the fact that all representations of the groups $E_8$ and $E_7$ are real (see, e.g., \cite{Slansky:1981yr}). Nevertheless, GUTs based on $E_8$ are sometimes considered in the literature, see, e.g., \cite{Konshtein:1980km,Baaklini:1980uq,Baaklini:1980fv,Bars:1980mb,Koca:1981xd,Ong:1984ej,Buchmuller:1985rc,Thomas:1985be,Barr:1987pu,Mahapatra:1988gc,Adler:2002yg,Adler:2004uj,Camargo-Molina:2016yqm,Morais:2020ypd,
Aranda:2020noz,Morais:2020odg,Aranda:2021bvg}.

In this paper we will also investigate a possibility of constructing GUT based on the group $E_8$ with the symmetry breaking pattern\footnote{Note that due to the $U_1$ factors this chain is different from the one considered in \cite{Thomas:1985be}.}

\begin{equation}\label{Symmetry_Breaking}
E_8\to E_7\times U_1 \to E_6\times U_1 \to SO_{10}\times U_1 \to SU_5 \times U_1 \to SU_3 \times SU_2 \times U_1.
\end{equation}

\noindent
It is assumed that all $U_1$ groups are different and are not subgroups of simple groups from the previous step. In particular, we will not consider such possible options of the symmetry breaking when at a certain step $n$ in the group $G_n\times U_1$ the factor $U_1$ is completely broken, while the group $G_n$ is broken down to $G_{n+1}\times U_1$. (This occurs if vacuum expectation values are acquired by Higgs fields, one of which lies in the adjoint representation of $G_n$ and has a vanishing $U_1$ charge, and the second one is a singlet with respect to $G_n$ with a nontrivial $U_1$ charge.) The choice of the chain (\ref{Symmetry_Breaking}) is motivated by the analogy with the flipped $SU_5$ model \cite{Barr:1981qv,Antoniadis:1987dx,Campbell:1987eb,Ellis:1988tx}, so that we will call the considered theory ``the flipped $E_8$ model''. The main advantage of the considered symmetry breaking pattern is that it can be realized without involving $E_8$ representations higher than adjoint 248. (Note that the adjoint representation 248 of $E_8$ coincides with its fundamental representation. In other words, 248 is the lowest nontrivial representation of $E_8$.) It is interesting to note that ${\cal N}=4$ supersymmetric Yang--Mills theory (which is free from ultraviolet divergences \cite{Sohnius:1981sn,Grisaru:1982zh,Howe:1983sr,Mandelstam:1982cb,Brink:1982pd}) contains matter superfields only in the adjoint representation, so that the use of ${\cal N}=4$ $E_8$ supersymmetric Yang--Mills theory for the unification of electroweak and strong interactions seems not so impossible. However, the study of this possibility goes far beyond the scope of this paper. Here we will concentrate on the study of the gauge coupling unification (in the tree approximation). Namely, we would like to reveal if it is possible to obtain the $SU_3\times SU_2\times U_1$ theory with

\begin{equation}\label{Standard_Coupling_Unification}
\alpha_2=\alpha_3; \qquad \sin^2\theta_W=3/8
\end{equation}

\noindent
at very high energies after the symmetry breaking (\ref{Symmetry_Breaking}) assuming that only various parts of the representation 248 acquire vacuum expectation values responsible for it. Certainly, this implies that we deal with the supersymmetric theories, because only in this case such a unification of the gauge couplings occurs.

Note that we do not discuss here the dynamical mechanism of the symmetry breaking. In particular, this implies that we do not indicate exactly the field content of the resulting theory. Really, we can find the representations responsible for a certain symmetry breaking, but without the dynamical mechanism of the symmetry breaking it is impossible to reveal the number of (super)fields which acquire the corresponding vacuum expectation values.

The paper is organized as follows. In Sect. \ref{Section_Gamma} we recall how to construct $\Gamma$-matrices satisfying the condition $\{\Gamma_i,\Gamma_j\} = 2\delta_{ij}\cdot 1$ in diverse dimensions. These matrices will be needed for constructing explicit expressions for the generators of the groups under consideration and their commutation relations. Some useful notations are introduced in Sect. \ref{Section_Notations}. In Sect. \ref{Section_E8_To_E7} we  investigate how the $E_8$ symmetry is broken down up to $E_7\times U_1$ by the vacuum expectation value of the representation 248. Also we find the relations between the coupling constants of the original $E_8$ theory and of the resulting $E_7\times U_1$ theory. The next symmetry breaking $E_7\times U_1 \to E_6\times U_1$ is considered in Sect. \ref{Section_E7_To_E6}. Sect. \ref{Section_E6_To_SO10} is devoted to the symmetry breaking $E_6\times U_1\to SO_{10}\times U_1$. In particular, we demonstrate that there are two different options for it. They produce three different options for the symmetry breaking $SO_{10}\times U_1 \to SU_5\times U_1$ considered in Sect. \ref{Section_SO10_To_SU5}. Finally, in Sect. \ref{Section_SU5_TO_MSSM} we consider the symmetry breaking $SU_5\times U_1\to SU_3\times SU_2\times U_1$ which can occur in 6 different ways. We demonstrate that the only one of them leads to the standard unification conditions (\ref{Standard_Coupling_Unification}), and only this option produces all representation required to accommodate the MSSM superfields. For the other options we find the values of the coupling constants, the values of the Weinberg angle (at the classical level), and the branching rules of the $E_8$ representation 248 with respect to the low energy $SU_3\times SU_2\times U_1$ group. The results are briefly summarized in Conclusion.

Note that throughout the paper we actually deal with the Lie algebras and ignore the global structure of the groups under consideration.

\section{Gamma-matrices in diverse dimensions}
\hspace*{\parindent}\label{Section_Gamma}

In this section we specify our notation for $\Gamma$-matrices in an arbitrary dimension $D$ (and the Euclidean signature), because these matrices appear in the commutation relations of the exceptional group generators. As we have already mentioned, these matrices satisfy the main condition

\begin{equation}
\{\Gamma_i,\Gamma_j\} = 2\delta_{ij}\cdot 1,
\end{equation}

\noindent
where $1$ is the identity matrix of the same size as $\Gamma_i$. It is well known that for even values of $D$ this size is $2^{D/2}\times 2^{D/2}$, while for odd values of $D$ the size of $\Gamma$-matrices is $2^{(D-1)/2}\times 2^{(D-1)/2}$. In particular, in two and three dimensions, as $\Gamma$-matrices, one can take the Pauli matrices,

\begin{eqnarray}
&&\qquad\qquad\qquad\quad \Gamma_1^{(2)} = \sigma_1 = \left(
\begin{array}{cc}
0 & 1\\
1 & 0
\end{array}
\right);\qquad \Gamma_2^{(2)} = \sigma_2 = \left(
\begin{array}{cc}
0 & -i\\
i & 0
\end{array}
\right);\\
&& \Gamma_1^{(3)} = \sigma_1 = \left(
\begin{array}{cc}
0 & 1\\
1 & 0
\end{array}
\right);\qquad \Gamma_2^{(3)} = \sigma_2 = \left(
\begin{array}{cc}
0 & -i\\
i & 0
\end{array}
\right);\qquad \Gamma_3^{(3)} = \sigma_3 = \left(
\begin{array}{cc}
1 & 0\\
0 & -1
\end{array}
\right).\qquad
\end{eqnarray}

\noindent
Note that in our notation the dimension $D$ is indicated in the subscript inside the round brackets if necessary. (However, sometimes we will omit this subscript if the corresponding equation is written for an arbitrary dimension.)

For the other values of $D$ the $\Gamma$-matrices can be constructed with the help of mathematical induction. Namely, we assume that we have constructed them in a certain {\it odd} dimension $D$. Then in the {\it even} dimension $D+1$ the $\Gamma$-matrices are defined as

\begin{equation}
\Gamma^{(D+1)}_i = \left(
\begin{array}{cc}
0 & \Gamma^{(D)}_i\\
\Gamma^{(D)}_i & 0
\end{array}
\right),\quad i=1,\ldots,D;\qquad \Gamma^{(D+1)}_{D+1} = \left(
\begin{array}{cc}
0 & -i\\
i & 0
\end{array}
\right)
\end{equation}

\noindent
and have a size twice as large as in the previous (odd) dimension. Moreover, in this case there is a martix

\begin{equation}\label{Gamma_Last}
\Gamma^{(D+1)}_{D+2} = \left(
\begin{array}{cc}
1 & 0 \\
0 & -1
\end{array}
\right).
\end{equation}

\noindent
It is easy to verify that it anticommutes with the other $\Gamma$-matrices and its square is equal to the identity matrix,

\begin{equation}
\big\{\Gamma^{(D+1)}_{D+2}, \Gamma^{(D+1)}_{i}\big\} = 0,\quad i=1,\ldots D+1;\qquad \big(\Gamma^{(D+1)}_{D+2}\big)^2 =1.\vphantom{\Big(}
\end{equation}

\noindent
Therefore, in the next (odd) dimension $D+2$, as $\Gamma$-matrices, we can take the $\Gamma$-matrices from the previous (even) dimension, $\Gamma_i^{(D+2)}\equiv \Gamma_i^{(D+1)}$ (where $i=1,\ldots D+1$), and add to them the matrix $\Gamma^{(D+2)}_{D+2} \equiv \Gamma^{(D+1)}_{D+2}$. This completes the induction step.

Note that all above constructed $\Gamma$-matrices are Hermitian,

\begin{equation}
(\Gamma_i)^+ = \Gamma_i.
\end{equation}

\noindent
For even values of $i$ they are pure imaginary and antisymmetric, while for odd values of $i$ they are real and symmetric.

The charge conjugation matrix is defined as the product of all symmetric $\Gamma$-matrixes. Below we will consider only the {\it even} dimensions $D$. In this case the charge conjugation matrix is given by the equation

\begin{equation}\label{C_Definition}
C \equiv \Gamma_1 \Gamma_3 \ldots \Gamma_{D-1}
\end{equation}

\noindent
and satisfy the relations

\begin{equation}\label{C_Properties}
C\,\Gamma_i\, C^{-1} = - (-1)^{D/2}\, (\Gamma_i)^T;\qquad C^{-1} = C^+ = C^T = (-1)^{D(D-2)/8}\, C.
\end{equation}

Below we will also use the antisymmetrized products of $\Gamma$-matrices

\begin{equation}\label{Gamma_Products}
\Gamma_{i_1 i_2\ldots i_k} \equiv \frac{1}{k!}\Big(\Gamma_{i_1}\Gamma_{i_2}\ldots \Gamma_{i_k}\pm\text{permutations}\Big).
\end{equation}

\noindent
From Eq. (\ref{C_Properties}) one can see that these matrices multiplied by $C$ (or by $\Gamma_{D+1} C$) are either symmetric or antisymmetric, namely, (for even $D$)

\begin{eqnarray}\label{GammaC_Symmetry}
&& (\Gamma_{i_1 i_2\ldots i_k} C)^T = (-1)^{(D-2k)(D-2k-2)/8}\, \Gamma_{i_1 i_2\ldots i_k} C;\vphantom{\Big(}\nonumber\\
&& (\Gamma_{i_1 i_2\ldots i_k} \Gamma_{D+1} C)^T = (-1)^{(D-2k)(D-2k+2)/8}\, \Gamma_{i_1 i_2\ldots i_k} \Gamma_{D+1} C.
\end{eqnarray}

\section{Notations}
\hspace*{\parindent}\label{Section_Notations}

In our notation the generators of the fundamental representation are denoted by $t_{\bm{A}}$, where $\bm{A}=1,\ldots,\text{dim}\, G$. Note that here we indicate the indices of the generators by bold letters because it is often convenient to present them as sets of certain subindices (which will be denoted by usual letters). We will always assume that the generators are normalized by the condition

\begin{equation}\label{Metric_Definition}
\mbox{tr} (t_{\bm{A}} t_{\bm{B}}) = \frac{1}{2} g_{\bm{AB}},
\end{equation}

\noindent
where $g_{\bm{AB}} = g_{\bm{BA}}$ is a metric. For the groups $SO_n$ or $SU_n$ it is convenient to choose it equal to $\delta_{\bm{AB}}$. However, for the groups $E_8$, $E_7$, and $E_6$ it is convenient to use a different metric which will be specified below. The matrix inverse to $g_{\bm{AB}}$ will be denoted by $g^{\bm{AB}}$.

The generators of an arbitrary representation $R$ denoted by $T_{\bm{A}}$ satisfy the relations

\begin{equation}\label{Casimir_Definitions}
\mbox{tr} (T_{\bm{A}} T_{\bm{B}}) = T(R) g_{\bm{AB}};\qquad  g^{\bm{AB}} (T_{\bm{A}} T_{\bm{B}})_{\bm{i}}{}^{\bm{j}} = C(R)_{\bm{i}}{}^{\bm{j}};\qquad [T_{\bm{A}}, T_{\bm{B}}] = i f_{\bm{AB}}{}^{\bm{C}} T_{\bm{C}},
\end{equation}

\noindent
where $f_{\bm{AB}}{}^{\bm{C}}$ are the structure constants. It is easy to see that the expression $f_{\bm{ABC}}\equiv g_{\bm{CD}} f_{\bm{AB}}{}^{\bm{D}}$ is antisymmetric with respect to the permutations of all indices. The generators of the adjoint representation and the group Casimir $C_2(G)$ can be constructed from the structure constants as

\begin{equation}
(T_{Adj\,\bm{A}})^{\bm{C}}{}_{\bm{B}} = if_{\bm{AB}}{}^{\bm{C}};\qquad C_2\, g_{\bm{AB}} \equiv - f_{\bm{AC}}{}^{\bm{D}} f_{\bm{BD}}{}^{\bm{C}}.
\end{equation}

\noindent
This in particular implies that

\begin{equation}\label{Double_Commutator}
g^{\bm{AB}} [T_{\bm{A}}, [T_{\bm{B}}, T_{\bm{C}}]] = C_2 T_{\bm{C}}.
\end{equation}

\noindent
For the irreducible representations

\begin{equation}\label{C(R)}
C(R)_{\bm{i}}{}^{\bm{j}} = C(R)\,\delta_{\bm{i}}{}^{\bm{j}},\quad \mbox{where}\quad C(R) = T(R)\cdot \frac{\mbox{dim}\,G}{\mbox{dim}\,R}.
\end{equation}

\noindent
(The last equation can easily be derived from Eq. (\ref{Casimir_Definitions}) by multiplying the first equation by $g^{\bm{AB}}$ and taking into account that $g^{\bm{AB}} g_{\bm{AB}}=\mbox{dim}\, G$ and $\delta_{\bm{i}}{}^{\bm{i}} = \mbox{dim}\, R$.)

\section{The symmetry breaking $E_8\to E_7\times U_1$}
\label{Section_E8_To_E7}

\subsection{The group $E_8$}
\hspace*{\parindent}\label{Subsection_E8}

Our starting point is a theory with the $E_8$ symmetry and a certain set of (super)fields in the adjoint representation 248. Note that 248 is a minimal nontrivial representation of $E_8$, so that in this case the adjoint and fundamental representations coincide. Following \cite{Green:1987sp}, to describe the group $E_8$, we will use the fact (see, e.g., \cite{Slansky:1981yr}) that it contains a subgroup $SO_{16}\subset E_8$. The corresponding branching rule for representation 248 is written as

\begin{equation}\label{248_Branching_Rule}
248\Big|_{E_8} = 120+128\Big|_{SO_{16}}.
\end{equation}

\noindent
Here 120 is the adjoint representation of $SO_{16}$, and 128 is its representation by Majorana--Weyl spinors. For definiteness, we will assume that 128 corresponds to the right spinors. Their indices will be indicated by the letters $a,b,$ etc. Note that the $\Gamma$-matrices in $D=16$ have the dimension $2^{16/2}\times 2^{16/2} = 256\times 256$, so that the spinors have 256 components. However, from Eq. (\ref{Gamma_Last}) we see that only first 128 components of the right spinor (which is obtained by acting $(1+\Gamma^{(16)}_{17})/2$ on an arbitrary 256-component spinor) do not vanish. Therefore, it is possible to assume that the spinor indices range from 1 to 128. Taking this into account from the branching rule (\ref{248_Branching_Rule}) we conclude that the generators of the group $E_8$ can be presented as the set

\begin{equation}
t_{\bm{A}} = \big\{t_a,\, t_{ij}\big\},
\end{equation}

\noindent
where the vector indices $i,j$ of the generators $t_{ij} = -t_{ji}$ range from 1 to 16 and the spinor index $a$ of $t_a$ ranges from 1 to 128. The commutators of these generators can be obtained using the explicit form of the $SO_{16}$ generators in the adjoint and spinor representations, symmetry properties, $SO_{16}$ tensor structure of various expressions, and the Jacobi identity. The normalization of the generators can be found with the help of Eq. (\ref{Double_Commutator}) in which it is necessary to set $C_2=1/2$ because the adjoint representation of $E_8$ coincides with the fundamental one. Writing the commutators we take into account that the matrix elements $\big(\Gamma^{(16)}_{ij} (1+\Gamma^{(16)}_{17})/2\big)_{ab}$ do not vanish only if $a,b=1,\ldots,128$ and in this case coincide with $\big(\Gamma^{(16)}_{ij}\big)_{ab}$. The result for the commutators of the $E_8$ generators can be presented in the form

\begin{equation}\label{E8_Commutators}
E_8\ \ \left\{
\begin{array}{l}
{\displaystyle [t_{ij},t_{kl}] = \frac{i}{\sqrt{120}}\Big(\delta_{il} t_{jk} - \delta_{jl} t_{ik} - \delta_{ik} t_{jl} + \delta_{jk} t_{il}\Big);\vphantom{\int\limits_p}}\\
{\displaystyle [t_{ij},t_a] = -\frac{i}{\sqrt{480}} (\Gamma^{(16)}_{ij})_a{}^b t_b;}\\
{\displaystyle [t_a,t_b] = - \frac{i}{2\sqrt{480}} (\Gamma^{(16)}_{ij} C^{(16)})_{ab} t_{ij},\vphantom{\int\limits^d}}
\end{array}
\right.
\end{equation}

\noindent
where the charge conjugation matrix $C^{(16)}$ is defined by Eq. (\ref{C_Definition}). According to Eqs. (\ref{C_Properties}) and (\ref{GammaC_Symmetry}), it is symmetric, while the matrices $\Gamma^{(16)}_{ij} C^{(16)}$ are antisymmetric. The corresponding metric (defined by Eq. (\ref{Metric_Definition})) is

\begin{equation}\label{E8_Metric}
g_{\bm{AB}} \to \left(
\begin{array}{cc}
{\displaystyle \delta_{ik}\delta_{jl}-\delta_{il}\delta_{jk}} & 0\\
0 & (C^{(16)})_{ab}
\end{array}
\right);\qquad g^{\bm{AB}} \to \left(
\begin{array}{cc}
{\displaystyle \frac{1}{4}\big(\delta_{ik}\delta_{jl}-\delta_{il}\delta_{jk}\big)} & 0\\
0 & (C^{(16)})^{ab}
\end{array}
\right).
\end{equation}

\noindent
Really, using Eq. (\ref{E8_Commutators}) one can verify that the above $E_8$ generators satisfy the equation

\begin{equation}
g^{\bm{AB}} t_{\bm{A}} t_{\bm{B}} = \frac{1}{2} t_{ij} t_{ij} + (C^{(16)})^{ab} t_a t_b = \frac{1}{2}.
\end{equation}

\noindent
Certainly, this identity is equivalent to the equations

\begin{eqnarray}
&& \frac{1}{2}[t_{ij},[t_{ij},t_{kl}]] + (C^{(16)})^{ab} [t_a,[t_b,t_{kl}]] = \frac{1}{2} t_{kl};\nonumber\\
&& \frac{1}{2}[t_{ij},[t_{ij},t_{d}]] + (C^{(16)})^{ab} [t_a,[t_b,t_{d}]] = \frac{1}{2} t_{d},
\end{eqnarray}

\noindent
which can easily be checked.

\subsection{The group $E_7$}
\hspace*{\parindent}\label{Subsection_E7}

The group $E_7$ can be described similarly to the $E_8$ case. For this purpose we will use the maximal subgroup $SO_{12}\times SO_3 \subset E_7$. Two lowest (nontrivial) irreducible representations of $E_7$ are the fundamental representation 56 and the adjoint representation 133 with the branching rules \cite{Slansky:1981yr}

\begin{eqnarray}\label{E7_Branching_Rules}
&& 56\Big|_{E_7} = [12,2] + [32,1]\Big|_{SO_{12}\times SO_3};\nonumber\\
&& 133\Big|_{E_7} = [1,3] + [32',2] + [66,1]\Big|_{SO_{12}\times SO_3}.
\end{eqnarray}

\noindent
Here $3$ and $66$ are the adjoint representations of $SO_3$ and $SO_{12}$, respectively, $32$ and $32'$ are two spinor representations of $SO_{12}$ corresponding to the Weyl spinors, and $2$ is the spinor representation of $SO_3$. Note that in $D=12$ the $\Gamma$-matrices have the size $2^{12/2}\times 2^{12/2} = 64\times 64$, but we will again use indices that range over half the values as in the $E_8$ case. They will be indicated by capital letters and range from 1 to 32. The indices corresponding to the left spinors will also be indicated by a dot. For definiteness, we will assume that the representations $32$ and $32'$ correspond to the right and left spinors, respectively. Then from the second equation in (\ref{E7_Branching_Rules}) we conclude that the $E_7$ generators are given by the set

\begin{equation}
t_{\bm{A}} = \big\{t_{ij}, t_\alpha, t_{a\dot{A}} \big\},
\end{equation}

\noindent
where the $SO_3$ spinor indices denoted by letters $a,b,\ldots$ take the values 1 and 2, and the $SO_3$ vector indices denoted by letters $\alpha,\beta,\ldots$ range from 1 to 3. The $SO_{12}$ vector indices $i,j,\ldots$ range from 1 to 12.

The commutation relations for the $E_7$ generators can again be obtained using the form of $SO_{3}$ and $SO_{12}$ generators in various representations, symmetry considerations, and the Jacobi identity. The result for the generators normalized by the condition (\ref{Metric_Definition}) can be written as

\begin{equation}\label{E7_Commutators}
E_7\ \ \left\{
\begin{array}{l}
{\displaystyle [t_\alpha,t_\beta] = \frac{i}{\sqrt{12}} \varepsilon_{\alpha\beta\gamma} t_\gamma;\qquad [t_\alpha,t_{ij}] = 0;\vphantom{\int\limits_p}}\\
{\displaystyle [t_\alpha,t_{a\dot{A}}] = -\frac{1}{2\sqrt{12}} (\sigma_\alpha)_a{}^b t_{b\dot{A}};\qquad [t_{ij}, t_{a\dot{A}}] = - \frac{i}{2\sqrt{24}} (\Gamma^{(12)}_{ij})_{\dot{A}}{}^{\dot{B}} t_{a\dot{B}};\vphantom{\int\limits_p}}\\
{\displaystyle [t_{ij},t_{kl}] = \frac{i}{\sqrt{24}}\Big(\delta_{il} t_{jk} - \delta_{jl} t_{ik} - \delta_{ik} t_{jl} + \delta_{jk} t_{il}\Big);}\\
{\displaystyle [t_{a\dot{A}}, t_{b\dot{B}}] = \frac{i}{4\sqrt{24}} (\sigma_2)_{ab} (\Gamma^{(12)}_{ij} C^{(12)})_{\dot{A}\dot{B}} t_{ij} + \frac{1}{2\sqrt{12}} (C^{(12)})_{\dot{A}\dot{B}} (\sigma_\alpha\sigma_2)_{ab} t_\alpha,\vphantom{\int\limits^d}}
\end{array}
\right.
\end{equation}

\noindent
where the metric has the form

$$
g_{\bm{AB}} \to \left(
\begin{array}{ccc}
{\displaystyle \delta_{ik}\delta_{jl}-\delta_{il}\delta_{jk}} & 0 & 0\\
0 & \delta_{\alpha\beta} & 0\\
0 & 0 & -(\sigma_2)_{ab} (C^{(12)})_{\dot{A}\dot{B}}
\end{array}
\right);
$$

\begin{equation}\label{E7_Metric}
g^{\bm{AB}} \to \left(
\begin{array}{ccc}
{\displaystyle \frac{1}{4}\big(\delta_{ik}\delta_{jl}-\delta_{il}\delta_{jk}\big)} & 0 & 0\\
0 & \delta_{\alpha\beta} & 0\\
0 & 0 & (\sigma_2)^{ab} (C^{(12)})^{\dot{A}\dot{B}}
\end{array}
\right).
\end{equation}

\noindent
The Pauli matrix $\sigma_2 = i\sigma_1\sigma_3$ is actually the charge conjugation matrix in $D=3$ multiplied by $i$. It is antisymmetric, while the matrices $\sigma_\alpha\sigma_2$ are symmetric. According to Eqs. (\ref{C_Properties}) and (\ref{GammaC_Symmetry}), the $D=12$ charge conjugation matrix $C^{(12)}$ is antisymmetric, while the matrices $\Gamma^{(12)}_{ij} C^{(12)}$ are symmetric,

\begin{equation}
(C^{(12)})^T = - C^{(12)};\qquad (C^{(12)})^2=-1;\qquad (\Gamma^{(12)}_{ij} C^{(12)})^T = \Gamma^{(12)}_{ij} C^{(12)}.
\end{equation}

\noindent
In particular, this implies that the metric (\ref{E7_Metric}) is really symmetric, while the right hand side in the last line of Eq. (\ref{E7_Commutators}) is antisymmetric (with respect to the simultaneous permutation of both indices).

In the explicit form the generators of the fundamental representation 56 can be constructed with the help of the first branching rule in Eq. (\ref{E7_Branching_Rules}) and are written as

\begin{eqnarray}
&& t_{ij} = \frac{i}{\sqrt{24}}
\left(
\begin{array}{cc}
(\delta_{ik}\delta_{jl} - \delta_{il}\delta_{jk})\delta_a^b & 0\\
0 & {\displaystyle \frac{1}{2}}(\Gamma^{(12)}_{ij})_{A}{}^{B}
\end{array}
\right);\nonumber\\
&& t_\alpha = \frac{1}{2\sqrt{12}} \left(
\begin{array}{cc}
\delta_{kl} (\sigma_\alpha)_a{}^b & 0\\
0 & 0
\end{array}
\right);\\
&& t_{d\dot{D}} = \frac{i}{2\sqrt{12}} \left(
\begin{array}{cc}
0 & (\sigma_2)_{da} (\Gamma^{(12)}_k)_{\dot{D}}{}^{B}\\
(\Gamma^{(12)}_l C^{(12)})_{A\dot{D}} \delta_d^b & 0
\end{array}
\right).\qquad\nonumber
\end{eqnarray}

\noindent
In particular, it is easy to verify that

\begin{equation}
C(56) \equiv g^{\bm{AB}} t_{\bm{A}} t_{\bm{B}} = \frac{1}{2} t_{ij} t_{ij} + t_\alpha t_\alpha + (\sigma_2)^{ab} (C^{(12)})^{\dot{A}\dot{B}} t_{a\dot{A}} t_{b\dot{B}} = \frac{19}{16} = \frac{1}{2}\cdot \frac{133}{56}
\end{equation}

\noindent
in agreement with Eq. (\ref{C(R)}).

\subsection{Details of the symmetry breaking}
\hspace*{\parindent}\label{Subsection_E8_To_E7}

First, we need to investigate if it is possible to break the $E_8$ symmetry by a vacuum expectation value of a scalar field in the representation 248 because our purpose is to avoid the use higher $E_8$ representations. As we have already mentioned, the group $E_8$ has the maximal subgroup $SO_{16}$, while one of $SO_{16}$ maximal subgroups is $SO_{12}\times SO_3\times SO_{3}$. Also we note that $SO_{12} \times SO_{3}$ is the maximal subgroup of $E_7$ with the corresponding brunching rules (\ref{E7_Branching_Rules}). Therefore, we obtain the embedding

\begin{equation}\label{SO12_SO3^2_Subgroup}
E_8 \supset SO_{16} \supset  \underbrace{SO_{12}\times SO_3}_{\subset E_7}\times SO_{3}.
\end{equation}

\noindent
This implies that the representations which appear in the branching rule of 248 with respect to $SO_{12}\times SO_3\times SO_3$ can also be combined into $E_7\times SO_3 \subset E_8$ representations. The result can be written as

\begin{eqnarray}\label{E8_To_E7_Branching_Rule}
&& 248\Big|_{E_8} = 120 + 128\Big|_{SO_{16}} \nonumber\\
&& = \underbrace{[1,1,3]}_{[1,3]\vphantom{\Big|_{E_7\times SO_3}}} + \underbrace{[1,3,1] + [66,1,1] + [32',2,1]}_{+ [133,1]\vphantom{\Big|_{E_7\times SO_3}}} + \underbrace{[12,2,2] + [32,1,2]}_{+ [56,2]\Big|_{E_7\times SO_3}}\Big|_{SO_{12}\times SO_3\times SO_3},\qquad
\end{eqnarray}

\noindent
where $E_7\times SO_3 \subset E_8$ representations are indicated below the curly brackets.

The scalar field in the representation 248 can be presented in the form

\begin{equation}
\varphi = \varphi_{\bm{A}}\, g^{\bm{AB}} t_{\bm{B}} = \frac{1}{2} \varphi_{ij} t_{ij} + \varphi_a (C^{(16)})^{ab} t_b.
\end{equation}

\noindent
Let assume that the components $\varphi_{13,14}$ and $\varphi_{15,16}$ acquire the same vacuum expectation values,\footnote{Note that in this paper we do not consider the dynamical mechanism which allows to give the vacuum expectation values (\ref{248_vevs}) to the scalar field(s) in 248 and will discuss only the form of vevs which should provide the proper symmetry breaking.}

\begin{equation}\label{248_vevs}
\big(\varphi_{13,14}\big)_0 = \big(\varphi_{15,16}\big)_0 = v_8,
\end{equation}

\noindent
and find the corresponding little group. By definition, vacuum expectation values of scalar fields should be invariant under the gauge transformations of the little group. Therefore, we need to find all $E_8$ generators which commute with $\varphi_0 = v_8 \big(t_{13,14} + t_{15,16}\big)$. Let us start with the generators $t_{ij}$. Evidently, they commute with $\varphi_0$ if $i,j=1,\ldots,12$. These generators correspond to the subgroup $SO_{12}$. It is easy to see that the generators of the little group also include

\begin{eqnarray}\label{E7_Little_Group_Even_Generators}
&& \widetilde t_1 \equiv \frac{1}{\sqrt{2}} \big(t_{13,16} - t_{14,15}\big); \qquad \widetilde t_2 \equiv \frac{1}{\sqrt{2}} \big(-t_{13,15} - t_{14,16}\big);\qquad\nonumber\\
&& \widetilde t_3 \equiv \frac{1}{\sqrt{2}} \big(t_{13,14} - t_{15,16}\big); \qquad \widetilde t_3' \equiv \frac{1}{\sqrt{2}}\big(-t_{13,14} - t_{15,16}\big).\qquad
\end{eqnarray}

\noindent
They are normalized in the standard way, $\mbox{tr}\big(\widetilde t_\alpha \widetilde t_\beta\big)=\delta_{\alpha\beta}/2$, $\mbox{tr}\big((\widetilde t_3')^2\big)=1/2$, $\mbox{tr}(\widetilde t_\alpha \widetilde t_3') = 0$ and generate the $SO_3\times U_1$ subgroup of the little group,

\begin{equation}\label{Widetilde_T_Commutators}
[\widetilde t_3', \widetilde t_\alpha] = 0;\qquad [\widetilde t_\alpha,\widetilde t_\beta] = \frac{i}{\sqrt{60}} \varepsilon_{\alpha\beta\gamma} \widetilde t_\gamma.
\end{equation}

\noindent
The $SO_3$ subgroup obtained in this way can be identified with the first $SO_3$ factor in the $SO_{12}\times SO_3\times SO_3$ subgroup (\ref{SO12_SO3^2_Subgroup}), while $\widetilde t_3'$ generates the $U_1$ subgroup of the second $SO_3$ factor. This second $SO_3$ subgroup is generated by the operators $\widetilde t_\alpha'$, where $\alpha = 1,2,3$ and

\begin{equation}
\widetilde t_1'\equiv \frac{1}{\sqrt{2}} \big(-t_{13,16} - t_{14,15}\big);\qquad \widetilde t_2' \equiv \frac{1}{\sqrt{2}} \big(t_{13,15} - t_{14,16}\big),
\end{equation}

\noindent
which satisfy the commutation relations

\begin{equation}\label{E8_SO3_Commutation_Relation}
[\widetilde t_\alpha', \widetilde t_\beta] = 0;\qquad [\widetilde t_\alpha', \widetilde t_\beta'] = \frac{i}{\sqrt{60}} \varepsilon_{\alpha\beta\gamma} \widetilde t_\gamma'.
\end{equation}

However, the little group is wider than $SO_{12}\times SO_3\times U_1$, because some of the $E_8$ generators $t_a$ also give 0 acting on the vacuum expectation value of the scalar field. Really, commuting $t_a$ and $\varphi_0$ with the help of Eq. (\ref{E8_Commutators}) we obtain

\begin{equation}\label{Little_Group_For_Spinors}
[\varphi_0, t_a] = v_8\, \big[t_{13,14}+t_{15,16},t_a\big] = -\frac{i v_8}{2\sqrt{120}} \big(\Gamma^{(16)}_{13,14} + \Gamma^{(16)}_{15,16}\big)_a{}^b\, t_b.
\end{equation}

\noindent
Using the explicit form of the $\Gamma$-matrices described in Sect. \ref{Section_Gamma} it is easy to check that the matrix present in this equation can be written as

\begin{equation}\label{Gamma_Sum}
-\frac{i}{2}\Big(\Gamma^{(16)}_{13,14} + \Gamma^{(16)}_{15,16}\Big) =
\left(
\begin{array}{cccc}
1 & 0 & 0 & 0\\
0 & 0 & 0 & 0\\
0 & 0 & -1 & 0\\
0 & 0 & 0 & 0
\end{array}
\right)
\cdot 1_{32} = \left(
\begin{array}{cc}
1 & 0 \\
0 & -1
\end{array}
\right) \cdot \frac{1}{2}\Big(1+\Gamma^{(12)}_{13}\Big).
\end{equation}

\noindent
Note that, as we have already discussed, the indices $a$ and $b$ in Eq. (\ref{Little_Group_For_Spinors}) range from 1 to 128 (although the size of $\Gamma$-matrices in $D=16$ is $256\times 256$), so that the matrix (\ref{Gamma_Sum}) has the size $128\times 128$.  In the last equality of Eq. (\ref{Gamma_Sum}) we also took into account that

\begin{equation}
\Gamma^{(12)}_{13} = \left(
\begin{array}{cc}
1 & 0\\
0 & -1
\end{array}
\right)\cdot 1_{32}.
\end{equation}

\noindent
Therefore, from Eq. (\ref{Little_Group_For_Spinors}) we conclude that the $E_8$ generators $t_a$ which belong to the little group form two left $SO_{12}$ spinors (each of them having 32 nontrivial components). Also it is easy to see that they belong to the spinor representation 2 of the group $SO_3$ with the generators $\widetilde t_\alpha$ defined in Eq. (\ref{E7_Little_Group_Even_Generators}), because in the spinor representation 128 of the group $SO_{16}$ they take the form

\begin{equation}
\widetilde T_\alpha = T(\widetilde t_\alpha) = \frac{1}{2\sqrt{60}}\, \sigma_\alpha \cdot \frac{1}{2}\Big(1-\Gamma^{(12)}_{13}\Big).
\end{equation}

\noindent
This implies that the spinor generators of the little group belong to the representation $[32',2]$ of the group $SO_{12}\times SO_3$. Using this fact from the second branching rule in Eq. (\ref{E7_Branching_Rules}) we see that for the vacuum expectation value (\ref{248_vevs}) the generators of the little group coincide with the generators of the subgroup $E_7\times U_1$. Thus, the vacuum expectation value (\ref{248_vevs}) really provides the required symmetry breaking

\begin{equation}
E_8\to E_7\times U_1.
\end{equation}

Our next purpose is to relate the values of the coupling constants for the little group $E_7\times U_1$ to the original $E_8$ coupling constant $e_8$. To do this, we first compare the commutation relations for the generators $t_{ij}$ in Eqs. (\ref{E8_Commutators}) and (\ref{E7_Commutators}). From these equations we conclude that the (properly normalized) generators $t_{ij}$ of $E_8$ and $E_7$ are related by the equation

\begin{equation}\label{E8_E7_Generator_Relation}
t_{ij}\Big|_{E_8} = \frac{1}{\sqrt{5}} t_{ij}\Big|_{E_7}.
\end{equation}

\noindent
Equivalently, it is possible to compare the commutation relations for the $E_8$ generators $\widetilde t_\alpha$ defined by Eq. (\ref{E7_Little_Group_Even_Generators}) and the $E_7$ generators $t_\alpha$. Then  from (the first line of) Eq. (\ref{E7_Commutators}) and Eq. (\ref{Widetilde_T_Commutators}) we obtain

\begin{equation}
\widetilde t_\alpha = \frac{1}{\sqrt{5}} t_\alpha,
\end{equation}

\noindent
and the numerical factor is really exactly the same as in Eq. (\ref{E8_E7_Generator_Relation}).

Standardly, the Yang--Mills field $A_\mu$ is presented in the form $A_\mu = i e A_\mu^{\bm{A}} t_{\bm{A}}$, where $e$ (or, equivalently, $\alpha=e^2/4\pi$) is the coupling constant. We always assume that the quadratic part of the Yang--Mills Lagrangian is written as

\begin{equation}
{\cal L} \to -\frac{1}{4} g_{\bm{AB}} \Big(\partial_\mu A_\nu^{\bm{A}} - \partial_\nu A_\mu^{\bm{A}} \Big) \Big(\partial_\mu A_\nu^{\bm{B}} - \partial_\nu A_\mu^{\bm{B}} \Big),
\end{equation}

\noindent
and the $t_{ij}$ generators (of the fundamental representation) of various group are normalized by the condition

\begin{equation}
\mbox{tr}(t_{ij} t_{kl}) = \frac{1}{2}\big(\delta_{ik}\delta_{jl} - \delta_{il}\delta_{jk}\big),
\end{equation}

\noindent
see, e.g., Eqs. (\ref{E8_Metric}) and (\ref{E7_Metric}). This implies that if generators differ by a numeric factor, the corresponding couplings $e$ should differ by an inverse numerical factor. Therefore, from Eq. (\ref{E8_E7_Generator_Relation}) we conclude that

\begin{equation}\label{E8_E7_Couplings}
e_7 = \frac{e_8}{\sqrt{5}},
\end{equation}

\noindent
where $e_8$ and $e_7$ are the coupling constants for the original $E_8$ and residual $E_7\times U_1$ theories, respectively. Certainly, $e_7$ is the coupling constant corresponding to the $E_7$ factor in the little group, and in this theory there is also the coupling $e^{(7)}_1$ which corresponds to the factor $U_1$.

The generator of the $U_1$ component of the little group $E_7\times U_1$ normalized in the same way as the $E_8$ generators is $\widetilde t_3'$ defined by Eq. (\ref{E7_Little_Group_Even_Generators}). However, it is desirable that the $U_1$ charges will be as simple as possible. That is why it is reasonable to require that they should be smallest possible integers. To find the corresponding generator, we first choose the generators of the $SO_3$ component of the group $E_7\times SO_3\subset E_8$ in the representations $2$ and $3$ in the form

\begin{equation}
t_\alpha'\Big|_{R=2} = \sigma_\alpha;\qquad
\big(t_\alpha'\big)_{\beta\gamma}\Big|_{R=3} = -2i\varepsilon_{\alpha\beta\gamma},
\end{equation}

\noindent
respectively. They satisfy the same commutation relations

\begin{equation}\label{Normalization_Structure_Constants}
[t_\alpha',t_\beta'] = 2i\varepsilon_{\alpha\beta\gamma} t_\gamma'.
\end{equation}

\noindent
The generator of the $U_1$ component of the little group evidently coincides with $t_3'$. Its eigenvalues in the representations $2$ and $3$ are $\pm 1$ and $-2,0,2$, respectively. Therefore, from Eq. (\ref{E8_To_E7_Branching_Rule}) we conclude that the branching rule for the $E_8$ adjoint representation $248$ with respect to the little group $E_7\times U_1$ reads as

\begin{equation}\label{E8_248_Branching_Rule}
248\Big|_{E_8} = 1(0) + 1(2) + 1(-2) + 133(0) + 56(1) + 56(-1)\Big|_{E_7\times U_1}.
\end{equation}

\noindent
We see that due to the normalization condition (\ref{Normalization_Structure_Constants}) the $U_1$ charges are really the least possible integers. However, the coupling constant $e_1^{(7)}$  corresponding to the $U_1$ subgroup is obviously different from $e_8$. To calculate it, we need to compare the commutation relation (\ref{Normalization_Structure_Constants}) with the commutation relation (\ref{E8_SO3_Commutation_Relation}) for the generators $\widetilde t_\alpha'$ normalized in the same way as all $E_8$ generators. From this comparison we conclude that

\begin{equation}
\widetilde t_3' = \frac{1}{4\sqrt{15}} t_3'.
\end{equation}

\noindent
Therefore, the corresponding coupling constants are related as

\begin{equation}\label{E7_U1_Coupling}
e_1^{(7)} = \frac{e_8}{4\sqrt{15}} = \frac{e_7}{4\sqrt{3}},
\end{equation}

\noindent
where for deriving the last equality we also used Eq. (\ref{E8_E7_Couplings}). Thus, we obtain that the coupling constants of the $E_7\times U_1$ theory are given by the expressions

\begin{equation}
\alpha_7 = \frac{\alpha_8}{5};\qquad \alpha_1^{(7)} = \frac{\alpha_7}{48},
\end{equation}

\noindent
where $\alpha_8\equiv e_8^2/4\pi$, $\alpha_7\equiv e_7^2/4\pi$, and $\alpha_1^{(7)} \equiv \big(e_1^{(7)}\big)^2/4\pi$.

\section{The symmetry breaking $E_7\times U_1 \to E_6\times U_1$}
\label{Section_E7_To_E6}

\subsection{The group $E_6$}
\hspace*{\parindent}\label{Subsection_E6}

The group $E_6$ can be described similarly to the groups $E_7$ and $E_8$. For this purpose we first note that $E_6$ has the maximal subgroup $SO_{10}\times U_1$. The corresponding branching rules for the lowest nontrivial representations are given by the equations \cite{Slansky:1981yr}

\begin{eqnarray}\label{E6_Branching_Rules}
&& 27\Big|_{E_6} = 1(4) + 10(-2)+16(1)\Big|_{SO_{10}\times U_1};\nonumber\\
&& \,\xbar{27}\,\Big|_{E_6} = 1(-4) + 10(2)+\,\xbar{16}\,(-1)\Big|_{SO_{10}\times U_1};\nonumber\\
&& 78\Big|_{E_6} = 1(0) + 16(-3) + \,\xbar{16}\,(3) + 45(0)\Big|_{SO_{10}\times U_1},
\end{eqnarray}

\noindent
where $27$ and $78$ are the fundamental and adjoint representations of $E_6$, respectively, and the $U_1$ charge is presented in the brackets. The $SO_{10}$ representations present in the right hand side can easily be constructed. Namely, the adjoint representation $45$ corresponds to the antisymmetric tensors with two indices, and the spinor representations $16$ and $\xbar{16}$ correspond to the right and left spinors. The size of the $\Gamma$-matrices in $D=10$ is $32\times 32$. In contrast to the $E_7$ case, here it is more convenient use a single spinor index which range from 1 to 32. As usual, the right spinors are the eigenvectors of the matrix $\Gamma^{(10)}_{11}$ with the eigenvalue 1, and all their components corresponding to $a=17,\ldots, 32$ vanish. Similarly, in the left spinors which correspond to the eigenvalue $(-1)$ the components with $a=1,\ldots,16$ vanish. Then, according to the third equation in (\ref{E6_Branching_Rules}), the $E_6$ generators can be presented as the set

\begin{equation}
t_{\bm{A}} = \big\{t_{ij}, t_a, t\big\},
\end{equation}

\noindent
where the vector indices $i,j$ range from 1 to 10. The generators $t_{ij}$ correspond to the representation $45$ of $SO_{10}$, the generators $t_a$ include $16$ and $\,\xbar{16}\,$, and $t$ is an $SO_{10}$ singlet. The commutation relations for $E_6$ can again be obtained from the branching rule for the representation $78$, symmetry considerations, and the Jacobi identity. The result can be written as

\begin{eqnarray}\label{E6_Commutators}
E_6\ \ \left\{
\begin{array}{l}
{\displaystyle [t_{ij},t_{kl}] = \frac{i}{\sqrt{12}}\Big(\delta_{il} t_{jk} - \delta_{jl} t_{ik} - \delta_{ik} t_{jl} + \delta_{jk} t_{il}\Big);\vphantom{\int\limits_p}}\\
{\displaystyle [t_{ij},t] = 0;\qquad [t,t_a] = \frac{1}{4} (\Gamma^{(10)}_{11})_a{}^b t_b;\qquad [t_{ij},t_a] = - \frac{i}{2\sqrt{12}} (\Gamma^{(10)}_{ij})_a{}^b t_b;}\\
{\displaystyle [t_a,t_b] = - \frac{i}{4\sqrt{12}} (\Gamma^{(10)}_{ij}C^{(10)})_{ab} t_{ij} + \frac{1}{4} (\Gamma^{(10)}_{11} C^{(10)})_{ab} t, \vphantom{\int\limits^d}}
\end{array}
\right.
\end{eqnarray}

\noindent
where the generators are normalized by Eq. (\ref{Metric_Definition}) with the metric

\begin{equation}\label{E6_Metric}
g_{\bm{AB}} \to \left(
\begin{array}{ccc}
{\displaystyle \delta_{ik}\delta_{jl}-\delta_{il}\delta_{jk}} & 0 & 0\\
0 & (C^{(10)})_{ab} & 0\\
0 & 0 & 1
\end{array}
\right);\qquad
g^{\bm{AB}} \to \left(
\begin{array}{ccc}
{\displaystyle \frac{1}{4}\big(\delta_{ik}\delta_{jl}-\delta_{il}\delta_{jk}\big)} & 0 & 0\\
0 & (C^{(10)})^{ab} & 0\\
0 & 0 & 1
\end{array}
\right).
\end{equation}

\noindent
Note that according to Eqs. (\ref{C_Properties}) and (\ref{GammaC_Symmetry}) the $D=10$ charge conjugation matrix $C^{(10)}$ is symmetric, while the matrices $\Gamma^{(10)}_{ij} C^{10}$ and $\Gamma^{(10)}_{11} C^{(10)}$ are antisymmetric,

\begin{eqnarray}
&& (C^{(10)})^T = C^{(10)};\qquad\qquad\qquad\quad (C^{10})^2 = 1; \qquad\vphantom{\Big(}\nonumber\\
&& (\Gamma^{(10)}_{ij} C^{(10)})^T = -\Gamma^{(10)}_{ij} C^{(10)};\qquad\, (\Gamma^{(10)}_{11} C^{(10)})^T = -\Gamma^{(10)}_{11} C^{(10)}.\qquad\vphantom{\Big(}
\end{eqnarray}

The generators of the fundamental representation $27$ can also be constructed explicitly with the help of the first branching rule in Eq. (\ref{E6_Branching_Rules}) and the commutation relations (\ref{E6_Commutators}). The result can be presented in the form

\begin{eqnarray}\label{E6_Generators_Of_27}
&&\hspace*{-5mm} t_{ij} = \frac{i}{\sqrt{12}} \left(
\begin{array}{ccc}
0 & 0 & 0\\
0 & \delta_{ik}\delta_{jl} - \delta_{il}\delta_{jk} & 0\vphantom{\Big(_a}\\
0 & 0 & {\displaystyle \frac{1}{4}} \left[\Gamma^{(10)}_{ij}(1+\Gamma^{(10)}_{11})\right]_a{}^b
\end{array}
\right);\nonumber\\
&&\hspace*{-5mm} t = \frac{1}{12}
\left(
\begin{array}{ccc}
4 & 0 & 0\vphantom{\Big(_a}\\
0 & -2\delta_{kl} & 0\\
0 & 0 & {\displaystyle \frac{1}{2}}(1+\Gamma^{(10)}_{11})_a{}^b
\end{array}
\right);\\
&&\hspace*{-5mm} t_d = \frac{1}{\sqrt{96}}
\left(
\begin{array}{ccc}
0 & 0 & \sqrt{2} \big(1+\Gamma^{(10)}_{11}\big)_d{}^b\vphantom{\Big(_a}\\
0 & 0 & \left[\Gamma^{(10)}_k(1+\Gamma^{(10)}_{11})\right]_d{}^b\\
\sqrt{2}\left[(1+\Gamma^{(10)}_{11}) C^{(10)}\right]_{ad} & \left[(1+\Gamma^{(10)}_{11}) \Gamma^{(10)}_l C^{(10)}\right]_{ad} & 0
\end{array}
\right).\nonumber
\end{eqnarray}

\noindent
As a correctness test, it is possible to check that

\begin{equation}
C(27) = g^{\bm{AB}} t_{\bm{A}} t_{\bm{B}} = \frac{1}{2} t_{ij} t_{ij} + (C^{(10)})^{ab} t_a t_b + t^2 = \frac{13}{9} = \frac{1}{2}\cdot \frac{78}{27}
\end{equation}

\noindent
in exact agreement with Eq. (\ref{C(R)}).

\subsection{The symmetry breaking}
\hspace*{\parindent}\label{Subsection_E7_To_E6}

According to \cite{Slansky:1981yr}, the group $E_7$ contains the maximal subgroup $E_6\times U_1$. The corresponding branching rules for the lowest $E_7$ representations are written as

\begin{eqnarray}\label{E7_Branching_Rules_To_E6}
&& 56\Big|_{E_7} = 27(1) + \xbar{27}\,(-1) + 1(3) + 1(-3)\Big|_{E_6\times U_1};\nonumber\\
&& 133\Big|_{E_7} = 1(0) + 27(-2) + \xbar{27}\,(2) + 78(0)\Big|_{E_6\times U_1}.
\end{eqnarray}

\noindent
Note that the $U_1$ charges (given in the brackets) were not presented in \cite{Slansky:1981yr}. For completeness, we describe their derivation in Appendix \ref{Appendix_E7_Branching}.

From Eq. (\ref{E7_Branching_Rules_To_E6}) we see that the representation $56$ contains two $E_6$ singlets with nontrivial $U_1$ charges. Therefore, if the scalar field corresponding to one of these singlets acquires a vacuum expectation value, then the little group will evidently contain a factor $E_6$. Certainly, the subgroup $U_1 \subset E_7$ is broken because the scalar field which acquires a vacuum expectation value has a nontrivial charge with respect to this group. However, we start with a theory invariant under $E_7\times U_1$ transformation, so that the gauge group contains a subgroup $E_6\times U_1 \times U_1$. From this $U_1\times U_1$ a certain $U_1$ subgroup survives in the little group. Really, let the representation $56(1)$ of $E_7\times U_1$ acquires the vacuum expectation value $v_7$. Without loss of generality we can assume that the corresponding scalar field belong to the representation $1(3)$ of $E_6\times U_1\subset E_7$. Let $\beta_2$ and $\beta_1$ are the parameters of two $U_1$ groups in $(E_6\times U_1)\times U_1 \subset E_7\times U_1$, respectively,

\begin{equation}\label{E6_U1^2_Subgroup}
E_7\times \underbrace{U_1}_{\beta_1^{(7)}} \supset (E_6\times \underbrace{U_1}_{\beta_2^{(7)}})\times \underbrace{U_1}_{\beta_1^{(7)}}.
\end{equation}

\noindent
Then the vacuum expectation value responsible for the symmetry breaking considered in this section is transformed as

\begin{equation}
v_7 \to \exp\Big(i\beta^{(7)}_1 + 3i\beta^{(7)}_2\Big)\, v_7.
\end{equation}

\noindent
This implies that it remains invariant under the transformations with $\beta^{(7)}_1 + 3\beta^{(7)}_2 = 0$, which obviously form the subgroup $U_1 \subset U_1\times U_1$.

It is important to find the relation between the coupling constants of the original $E_7\times U_1$ theory and its $E_6\times U_1$ remnant. The relation between the $E_7$ and $E_6$ couplings can be found by comparing the commutation relations for the (properly normalized) $t_{ij}$ generators of these groups. They are given by the third line in Eq. (\ref{E7_Commutators}) and by the first line in Eq. (\ref{E6_Commutators}), respectively. Requiring that after the symmetry breaking\footnote{The gauge fields corresponding to the $U_1$ groups will be considered separately.}

\begin{equation}
A_\mu\Big|_{E_7} = i e_7 A_\mu^{\bm{A}} t_{\bm{A}}\Big|_{E_7}\quad \to\quad A_\mu\Big|_{E_6} = ie_6 A_\mu^{\bm{A}} t_{\bm{A}}\Big|_{E_6}
\end{equation}

\noindent
and noting that $t_{ij}$ for $E_7$ are in $\sqrt{2}$ times smaller than the ones for $E_6$, we see that the coupling constants should be related by the equation

\begin{equation}\label{E7_E6_Coupling_Relation}
e_6 = \frac{e_7}{\sqrt{2}}.
\end{equation}

To find the coupling constant corresponding to the $U_1$ component of the little group $E_6\times U_1$, we first write the branching rule for the representation $56(1)$ with respect to the subgroup (\ref{E6_U1^2_Subgroup}) using the first equation in (\ref{E7_Branching_Rules_To_E6}),

\begin{equation}
56(1)\Big|_{E_7\times U_1} = 27(1,1) + \,\xbar{27}\,(1,-1) + \bm{1(1,3)} + 1(1,-3)\Big|_{E_6\times U_1\times U_1},
\end{equation}

\noindent
where the $E_6$ singlet which acquires a vacuum expectation value is indicated by the bold font. In the brackets we present the charges $Q_1^{(7)}$ and $Q_2^{(7)}$ corresponding to the $U_1$ subgroups parameterized by $\beta_1^{(7)}$ and $\beta_2^{(7)}$, respectively.\footnote{To avoid supersymmetry breaking by the $D$-term corresponding to the $U_1$ group, the part $1(-1,-3)$ of the representation $56(-1)$ should also acquire the same vacuum expectation value. However, in this paper we do not consider dynamics of the theory and will not discuss the corresponding issues.} This implies that under the $U_1\times U_1$ subgroup in (\ref{E6_U1^2_Subgroup}) a (super)field $\varphi$ changes as

\begin{equation}\label{E7_U1^2_Transformation}
\varphi \to \exp\Big(i Q_1^{(7)} \beta_1^{(7)} + i Q_2^{(7)} \beta_2^{(7)}\Big)\varphi.
\end{equation}

\noindent
Evidently, the $U_1$ charge corresponding to the little group can be chosen in the form

\begin{equation}\label{E6_Little_Group_Charge}
Q_1^{(6)} = \frac{1}{2} \Big(- 3 Q_1^{(7)} + Q_2^{(7)} \Big)
\end{equation}

\noindent
because in this case the $E_6$ singlet acquiring a vacuum expectation value will really have the required charge 0 with respect to the little group. The factor $1/2$ is included for the further convenience. Namely, due to this factor various terms in the branching rule for the $E_8$ representation 248 with respect to the $E_6\times U_1$ subgroup will have the least possible integer values of $U_1$ charges, see Eq. (\ref{248_To_E6}) below. The transformation of a (super)field under the little group can be written as

\begin{equation}\label{E7_U1_Little_Group}
\varphi \to \exp\Big(i Q_1^{(6)}\beta_1^{(6)} \Big)\varphi,
\end{equation}

\noindent
and our aim is to find the coupling constant $e_1^{(6)}$ corresponding to this group.

First, we note that, according to Eq. (\ref{E7_U1_Coupling}), the coupling constant for the $U_1$ component of $E_7\times U_1$ is given by the expression

\begin{equation}
e_1^{(7)} = \frac{e_7}{4\sqrt{3}}.
\end{equation}

\noindent
This implies that the charge $Q_1^{(7)}$ is an eigenvalue of the operator $4\sqrt{3}\,t_1^{(7)}$, where $t_1^{(7)}$ is the generator of the $U_1$ component in the subgroup $E_7\times U_1$ normalized in the same way as the generators of the group $E_7$.

Let us also introduce the generator $t\Big|_{U_1\subset E_7}$ of the $U_1$ component in the subgroup $E_6\times U_1 \subset E_7$ normalized in same way as the other $E_7$ generators. According to the first branching rule in Eq. (\ref{E7_Branching_Rules_To_E6}), it can presented as

\begin{equation}\label{T_2^7_Generator}
t\Big|_{U_1\subset E_7} =
\frac{1}{12} \left(
\begin{array}{cccc}
3 & 0 & 0 & 0\\
0 & -3 & 0 & 0\\
0 & 0 & 1 & 0\\
0 & 0 & 0 & -1
\end{array}
\right)\quad
\text{acting on} \quad
\left(
\begin{array}{c}
1\\
1\\
27\\
\xbar{27}
\end{array}
\right).
\end{equation}

\noindent
Note this generator is really properly normalized because

\begin{equation}
\mbox{tr}\left(\Big(t\Big|_{U_1\subset E_7}\Big)^2\right) = \frac{1}{144} \Big(1\cdot 3^2 + 1\cdot (-3)^2 + 27\cdot 1^2 + 27\cdot 1^2\Big) = \frac{72}{144} = \frac{1}{2}.
\end{equation}

\noindent
Comparing Eq. (\ref{T_2^7_Generator}) with the first branching rule in Eq. (\ref{E7_Branching_Rules_To_E6}) we see that the charge $Q_2^{(7)}$ can be considered as an eigenvalue of the operator $12\, t\Big|_{U_1\subset E_7}$. Therefore, the little group charge (\ref{E6_Little_Group_Charge}) is an eigenvalue of the operator

\begin{equation}
\frac{1}{2} \Big(- 3\cdot 4\sqrt{3}\, t_1^{(7)} + 12\, t\Big|_{U_1\subset E_7} \Big) = 12 \bigg(- \frac{\sqrt{3}}{2} t_1^{(7)} + \frac{1}{2} t\Big|_{U_1\subset E_7}\bigg).
\end{equation}

\noindent
The operator in the brackets in the right hand side of this equation is normalized in the same way as all generators of the group $E_7$. Therefore, the numerical coefficient before this operator indicates how many times the coupling constant for the $U_1$ component of little group is less than the $E_7$ coupling constant,

\begin{equation}
e_1^{(6)} = \frac{e_7}{12}.
\end{equation}

\noindent
Taking into account Eq. (\ref{E7_E6_Coupling_Relation}), which relates the couplings corresponding to the $E_6$ and $E_7$ groups, we obtain the values for the coupling constants for the resulting $E_6\times U_1$ theory,

\begin{equation}\label{E6_U1_Couplings}
e_6 = \frac{e_7}{\sqrt{2}};\qquad e_1^{(6)} = \frac{e_7}{12} = \frac{e_6}{6\sqrt{2}}.
\end{equation}

\noindent
This equation can equivalently be rewritten as

\begin{equation}\label{E6_Couplings}
\alpha_6 = \frac{\alpha_7}{2};\qquad \alpha_1^{(6)} = \frac{\alpha_6}{72}.
\end{equation}

Using the branching rules (\ref{E8_248_Branching_Rule}) and (\ref{E7_Branching_Rules_To_E6}) we obtain the decomposition of the $E_8$ representation $248$ with respect to the $E_6\times U_1\times U_1$ subgroup (\ref{E6_U1^2_Subgroup}),

\begin{eqnarray}\label{E8_E6_Preliminary}
&& 248\Big|_{E_8} = \Big[1(0,0) + 1(2,0) + 1(-2,0)\Big] + \Big[1(0,0) + 27(0,-2) + \,\xbar{27}\,(0,2) + 78(0,0)\Big] \nonumber\\
&&\qquad\quad\ + \Big[27(1,1) + \,\xbar{27}\,(1,-1) + \bm{1(1,3)} + 1(1,-3)\Big] + \Big[27(-1,1) + \,\xbar{27}\,(-1,-1) \nonumber\\
&&\qquad\quad\ + 1(-1,3) + 1(-1,-3)\Big]\bigg|_{E_6\times U_1\times U_1},
\end{eqnarray}

\noindent
where in the round brackets we present the charges $Q_1^{(7)}$ (the first number) and $Q_2^{(7)}$ (the second number). Calculating the little group charge (\ref{E6_Little_Group_Charge}) for each term in this expression we construct the branching rule for the $E_8$ representation $248$ with respect to little group $E_6\times U_1$,

\begin{eqnarray}\label{248_To_E6}
&& 248\Big|_{E_8} = \Big[1(0) + 1(-3) + 1(3)\Big] + \Big[1(0) + 27(-1) + \,\xbar{27}\,(1) + 78(0)\Big] \nonumber\\
&&\qquad\quad\ + \Big[27(-1) + \,\xbar{27}\,(-2) + \bm{1(0)} + 1(-3)\Big] + \Big[27(2) + \,\xbar{27}\,(1) + 1(3) + 1(0)\Big]\bigg|_{E_6\times U_1}\nonumber\\
&&\qquad\quad\ = 4\times 1(0) + 2\times 1(3) + 2\times 1(-3) + 2\times 27(-1) + 2\times\,\xbar{27}\,(1) + 27(2) \vphantom{\Big|_{E_6}}\nonumber\\
&&\qquad\quad\ + \,\xbar{27}\,(-2) + 78(0)\Big|_{E_6\times U_1}.
\end{eqnarray}

Below we will see that the further symmetry breaking (down to $SO_{10}\times U_1$) can be realized with the help of vacuum expectation values of the representations $27$ (and/or $\,\xbar{27}\,$) with nontrivial $U_1$ charges. From Eq. (\ref{248_To_E6}) we see that there are two different possible values for the $U_1$ charge of the representation $27$, so that the further symmetry breaking can be made by two different ways, which will be considered in the next section.

\section{The symmetry breaking $E_6\times U_1 \to SO_{10} \times U_1$}
\label{Section_E6_To_SO10}

\subsection{$SO_{10}$ coupling constant}
\hspace*{\parindent}

We will choose the $SO_{10}$ generators in the form

\begin{equation}
(t_{ij})_{kl} = \frac{i}{2}\left(\delta_{ik}\delta_{jl} - \delta_{il}\delta_{jk}\right).
\end{equation}

\noindent
In this case they satisfy the normalization condition (\ref{Metric_Definition}) with the metric

\begin{equation}
g_{\bm{AB}} \to \delta_{ik}\delta_{jl} - \delta_{il}\delta_{jk};\qquad g^{\bm{AB}} \to \frac{1}{4}\Big(\delta_{ik}\delta_{jl} - \delta_{il}\delta_{jk}\Big)
\end{equation}

\noindent
and the commutation relations

\begin{equation}
[t_{ij},t_{kl}] = \frac{i}{2}\Big(\delta_{il} t_{jk} - \delta_{jl} t_{ik} - \delta_{ik} t_{jl} + \delta_{jk} t_{il}\Big).
\end{equation}

Comparing this commutator with the commutator of the generators $t_{ij}$ for the group $E_6$ (presented in the first line of Eq. (\ref{E6_Commutators})) and requiring that the $E_6$ gauge field gives the $SO_{10}$ gauge field after the symmetry breaking,

\begin{equation}
A_\mu\Big|_{E_6} = i e_6 A_\mu^{\bm{A}} t_{\bm{A}}\Big|_{E_6}\quad \to\quad A_\mu\Big|_{SO_{10}} = ie_{10} A_\mu^{\bm{A}} t_{\bm{A}}\Big|_{SO_{10}} = \frac{i}{2} e_{10} \big(A_\mu\big)_{ij} t_{ij}\Big|_{SO_{10}},
\end{equation}

\noindent
we obtain the relation

\begin{equation}\label{SO10_Coupling}
e_{10} = \frac{e_6}{\sqrt{3}}.
\end{equation}

It is more complicated to find an analogous equation for the coupling constant of the $U_1$ subgroup. In fact, the form of this relation depends on the symmetry breaking pattern. According to Eq. (\ref{E6_Branching_Rules}), the representation $27$ of $E_6$ contains an $SO_{10}$ singlet. Certainly, so does the conjugated representation $\,\xbar{27}\,$. This implies that this representation can be used for breaking $E_6\times U_1$ down to $SO_{10}\times U_1$. However, the decomposition (\ref{248_To_E6}) of the $E_8$ representation $248$ contains the representations $27$ and $\,\xbar{27}\,$ with different $U_1$ charges. Evidently, there are only two essentially different options:

1. A vacuum expectation value is acquired by the representations $27(-1)$ and/or $\,\xbar{27}\,(1)$;

2. A vacuum expection value is acquired by the representations $27(2)$ and/or $\,\xbar{27}\,(-2)$.

\noindent
(Really, if a certain representation contains a singlet with respect to the little group, then the conjugated representation will also contain a singlet with respect to the same little group.)

Below we will separately consider both options for the symmetry breaking calling them B-1 and B-2, respectively. Note that here and below the first option always corresponds to a smaller absolute value of the corresponding $U_1$ charge, while the second option corresponds to a larger one.

\subsection{Symmetry breaking by the vacuum expectation value of $27(-1)$ (B-1)}
\hspace*{\parindent}

According to the first equation in Eq. (\ref{E6_Branching_Rules}), the $E_6$ representation $27$ contains a singlet with respect to the $SO_{10}$ factor in the subgroup $SO_{10}\times U_1\subset E_6$, which, however, has a nontrivial charge with respect to the $U_1$ factor. Nevertheless, the original theory has a symmetry $E_6\times U_1$, which is wider than $E_6$. Therefore, the little group will also contain a $U_1$ factor composed of two $U_1$ subgroups in the embedding

\begin{equation}\label{SO10_U1^2_Subgroup}
E_6\times \underbrace{U_1}_{\beta_1^{(6)}} \supset (SO_{10}\times \underbrace{U_1}_{\beta_2^{(6)}})\times \underbrace{U_1}_{\beta_1^{(6)}}.
\end{equation}

\noindent
The parameters of these $U_1$ groups denoted by $\beta_1^{(6)}$ and $\beta_2^{(6)}$ are indicated under each group. This means that a certain (super)field with the charges $Q_1^{(6)}$ and $Q_2^{(6)}$ under the $U_1\times U_1$ transformations changes as

\begin{equation}\label{E6_B1_U1^2_Transformation}
\varphi \to \exp\Big(i Q_1^{(6)} \beta_1^{(6)} + i Q_2^{(6)} \beta_2^{(6)}\Big)\varphi.
\end{equation}

\noindent
With respect to the subgroup (\ref{SO10_U1^2_Subgroup}) the branching rule for the $E_6\times U_1$ representation $27(-1)$ is written as

\begin{equation}\label{27_Branching_Rule}
27(-1)\Big|_{E_6\times U_1} = 16(-1,1) + 10(-1,-2) + \bm{1(-1,4)}\Big|_{SO_{10}\times U_1\times U_1},
\end{equation}

\noindent
where we present $(Q_1^{(6)},Q_2^{(6)})$ in the brackets and indicate the $SO_{10}$ singlet by the bold font. From this equation it is obvious that the little group contains a $U_1$ factor under which the transformation (\ref{E6_B1_U1^2_Transformation}) is reduced to

\begin{equation}\label{E6_U1_Little_Group}
\varphi \to \exp\Big(i Q_1^{(10)}\beta_1^{(10)} \Big)\varphi\qquad\mbox{with}\qquad Q_1^{(10)} \equiv \frac{1}{3} \Big(4 Q_1^{(6)} + Q_2^{(6)} \Big),
\end{equation}

\noindent
because with respect to this group the $SO_{10}$ singlet has the charge $Q_1^{(10)} = \frac{1}{3}(4\cdot (-1) + 4) = 0$. We would like to find the value of the coupling constant $e_1^{(10)}$ for this $U_1$ group.  According to Eq. (\ref{E6_U1_Couplings})

\begin{equation}
e_1^{(6)} = \frac{e_6}{6\sqrt{2}},
\end{equation}

\noindent
so that the expression $Q_1^{(6)}/6\sqrt{2}$ is the eigenvalue of the $U_1$ generator $t_1^{(6)}$ which is normalized in the same way as the generators of the group $E_6$. Therefore, the charge $Q_1^{(6)}$ is the eigenvalue of the operator $6\sqrt{2}\, t_1^{(6)}$. To construct the analogous expression for $Q_2^{(6)}$ we compare Eq. (\ref{27_Branching_Rule}) with the explicit expression for the $E_6$ generator $t = t\Big|_{U_1\subset E_6}$ in the fundamental representation given by Eq. (\ref{E6_Generators_Of_27}). Then we conclude that the charge $Q_2^{(6)}$ is an eigenvalue of the operator $12\, t$. Therefore, the value $(4Q_1^{(6)} + Q_2^{(6)})/3$ corresponds to the operator

\begin{equation}
\frac{1}{3} \Big(4\cdot 6\sqrt{2}\,t_1^{(6)} + 12 t\Big) = 12\bigg(\frac{\sqrt{8}}{3} t_1^{(6)} + \frac{1}{3} t\bigg).
\end{equation}

\noindent
In the right hand side the operator in the brackets is normalized in the same way as the $E_6$ generators, because

\begin{equation}
\Big(\frac{\sqrt{8}}{3}\Big)^2 + \Big(\frac{1}{3}\Big)^2 = 1.
\end{equation}

\noindent
Therefore, the coupling constant $e_1^{(10)}$ can be expressed in terms of the coupling constant $e_6$ as

\begin{equation}\label{SO10_U1_B1_Coupling}
e_1^{(10)} = \frac{e_6}{12} = \frac{e_{10}}{4\sqrt{3}},
\end{equation}

\noindent
where we also took Eq. (\ref{SO10_Coupling}) into account. Thus, we obtain that the coupling constants of the $SO_{10}\times U_1$ theory obtained after the symmetry breaking are related to each other and to the $E_6$ coupling constant by the equations

\begin{equation}\label{SO10_B1_Couplings}
\mbox{B-1:}\qquad \alpha_{10} = \frac{\alpha_6}{3};\qquad \alpha_1^{(10)} = \frac{\alpha_{10}}{48}.
\end{equation}

Next, we should construct a branching rule for the $E_8$ representation $248$ with respect to the $SO_{10}\times U_1$ subgroup. For this purpose we first write the branching rule with respect to the $SO_{10}\times U_1\times U_1$ subgroup (\ref{SO10_U1^2_Subgroup}) using Eqs. (\ref{E6_Branching_Rules}) and (\ref{248_To_E6}). The result can be written as

\begin{eqnarray}\label{E8_SO10_Preliminary}
&& 248\Big|_{E_8} = 5\times 1(0,0) + 2\times 1(3,0) + 2\times 1(-3,0) + 2\times \bm{1(-1,4)} + 2\times 1(1,-4)\qquad  \nonumber\\
&&\qquad\quad\ + 1(2,4) + 1(-2,-4) + 2\times 10(1,2) + 2\times 10(-1,-2) + 10(2,-2) \vphantom{\Big(}\nonumber\\
&&\qquad\quad\ + 10(-2,2)  + 16(0,-3) + \,\xbar{16}\,(0,3) + 2\times 16(-1,1) + 2\times\,\xbar{16}\,(1,-1)\vphantom{\Big(}\nonumber\\
&&\qquad\quad\ + 16(2,1) + \,\xbar{16}\,(-2,-1) + 45(0,0)\Big|_{SO_{10}\times U_1\times U_1}.
\end{eqnarray}

\noindent
Calculating the charges $Q_1^{(10)}$ with respect to the little group $SO_{10}\times U_1$ with the help of Eq. (\ref{E6_U1_Little_Group}) we obtain the required branching rule

\begin{eqnarray}\label{248_To_SO10_B1}
&& 248\Big|_{E_8} = 9\times 1(0) + 3\times 1(4) + 3\times 1(-4) +3\times 10(-2) + 3\times 10(2)\qquad\nonumber\\
&&\qquad\quad\ + 3\times 16(-1) + 3\times \,\xbar{16}\,(1) + 16(3) + \,\xbar{16}\,(-3) + 45(0)\Big|_{SO_{10}\times U_1}.
\end{eqnarray}

In Sect. \ref{Section_SO10_To_SU5} we will see that the symmetry breaking $SO_{10}\times U_1 \to SU_5\times U_1$ can be realized if a vacuum expectation value is acquired by a scalar field in the representation $16$ of the group $SO_{10}$. Again, in this case there are two different options for this symmetry breaking because this scalar field can belong either to representations $16(-1)$ (and/or $\,\xbar{16}\,(1)$) or $16(3)$ (and/or $\,\xbar{16}\,(-3)$). All these options (called B-1-1 and B-1-2, respectively) will be considered in Sect. \ref{Section_SO10_To_SU5}.

\subsection{Symmetry breaking by the vacuum expectation value of $27(2)$ (B-2)}
\hspace*{\parindent}

Alternatively, the symmetry breaking $E_6\times U_1 \to SO_{10}\times U_1$ can be realized if a vacuum expectation value is acquired by a scalar field in the represetation $27(2)$ (and/or $\,\xbar{27}\,(-2)$) of $E_6\times U_1$. Although in this case the little group is the same as in the option B-1, the $U_1$ coupling constant (as we will see below) will be different.

Taking into account that the branching rule for the representation $27(2)$ with respect to the $SO_{10}\times U_1\times U_1$ subgroup (\ref{SO10_U1^2_Subgroup}) is written as

\begin{equation}\label{27_B2_Branching_Rule}
27(2)\Big|_{E_6\times U_1} = 16(2,1) + 10(2,-2) + \bm{1(2,4)}\Big|_{SO_{10}\times U_1\times U_1},
\end{equation}

\noindent
we see that the charge with respect to the $U_1$ component of the little group is proportional to $2Q_1^{(6)} - Q_2^{(6)}$. Really, in this case the charge of the $SO_{10}$ singlet present in the representation $27(2)$ (which is indicated by the bold font in Eq. (\ref{27_B2_Branching_Rule})) vanishes. It is convenient to choose the normalization constant equal to $1/3$, so that for the option B-2 the charge with respect to the $U_1$ component of the little group is defined as

\begin{equation}\label{SO10_B2_Little_Group_Charge}
Q_1^{(10)} = \frac{1}{3} \Big(2 Q_1^{(6)} - Q_2^{(6)} \Big).
\end{equation}

\noindent
Exactly as for the option B-1, we see that this charge is an eigenvalue of the operator

\begin{equation}\label{E6_To_SO10_B2_Normalized_Generator}
\frac{1}{3} \Big(2\cdot 6\sqrt{2}\, t_1^{(6)} - 12 t\Big) = 4\sqrt{3}\bigg(\sqrt{\frac{2}{3}}\, t_1^{(6)} - \sqrt{\frac{1}{3}}\, t\bigg).
\end{equation}

\noindent
Again, in the right hand side the operator in the brackets is normalized in same way as all $E_6$ generators because

\begin{equation}
\Big(\sqrt{\frac{2}{3}}\,\Big)^2 + \Big(-\sqrt{\frac{1}{3}}\,\Big)^2 = 1.
\end{equation}

\noindent
From Eq. (\ref{E6_To_SO10_B2_Normalized_Generator}) we conclude that the coupling constant for the $U_1$ component of the little group is related to the $E_6$ coupling constant by the equation

\begin{equation}\label{SO10_U1_B2_Coupling}
e_1^{(10)} = \frac{e_6}{4\sqrt{3}} = \frac{e_{10}}{4},
\end{equation}

\noindent
where we also involved Eq. (\ref{SO10_Coupling}). Thus, we obtain that the couplings of the $SO_{10}\times U_1$ theory obtained after the symmetry breaking by a vacuum expectation value of $27(2)$ of $E_6\times U_1$ are given by the expressions

\begin{equation}\label{SO10_B2_Couplings}
\mbox{B-2:}\qquad \alpha_{10} = \frac{\alpha_6}{3};\qquad \alpha_1^{(10)} = \frac{\alpha_{10}}{16}.
\end{equation}

\noindent
In particular, we see that the value of the $U_1$ coupling constant is different from the one in the variant B-1 (see Eq. (\ref{SO10_B1_Couplings})). Therefore, these two options of the symmetry breaking are really essentially different.

Calculating the charges $Q_1^{(10)}$ for various terms in Eq. (\ref{E8_SO10_Preliminary}) with the help of Eq. (\ref{SO10_B2_Little_Group_Charge}) we derive the branching rule for the representation 248 of $E_8$ with respect to the little group $SO_{10}\times U_1$,

\begin{eqnarray}\label{248_To_SO10_B2}
&& 248\Big|_{E_8} = 7\times 1(0) + 4\times 1(2) + 4\times 1(-2) +4\times 10(0) + 10(2) + 10(-2)\nonumber\\
&&\qquad\quad\ + 2\times 16(1) + 2\times \,\xbar{16}\,(-1) + 2\times 16(-1)
+ 2\times \,\xbar{16}\,(1) + 45(0)\Big|_{SO_{10}\times U_1}.\qquad
\end{eqnarray}

Note that in this case vacuum expectation values of the representations $16(1)$ and $16(-1)$ are actually equivalent because the difference in the sign of the $U_1$ coupling constant is not essential. Therefore, in this case there is the only option for the further symmetry breaking, which will be called B-2-1 below.

\section{The symmetry breaking $SO_{10}\times U_1 \to SU_{5} \times U_1$}
\label{Section_SO10_To_SU5}

\subsection{The embedding $SU_5\times U_1 \subset SO_{10}$}
\hspace*{\parindent}\label{Subsection_SO10_To_SU5_Embedding}

To construct the embedding $U_5\subset SO_{10}$, we consider a complex 5-component column $z = x+iy$ in the fundamental representation of the group $U_5$. This implies that it transforms as

\begin{equation}\label{Z_Transformation}
z\equiv x+iy \to \Omega_5 z = (B+iC)(x+iy) = (Bx -Cy) + i(By + Cx),
\end{equation}

\noindent
where the $5\times 5$ matrix $\Omega_5 \in U_5$ was written as the sum of the real part $B$ and the purely imaginary part $iC$. Note that from the condition $\Omega_5^+ \Omega_5=1$ we obtain that the real matrices $B$ and $C$ satisfy the constraints

\begin{equation}\label{BC_Constraints}
B^T B + C^T C = 1;\qquad B^T C = C^T B.
\end{equation}

\noindent
The transformation (\ref{Z_Transformation}) can equivalently be presented as the transformation of a real 10-component column

\begin{equation}\label{U5_Subgroup}
\left(
\begin{array}{c}
x\\y
\end{array}
\right) \to
\left(
\begin{array}{cc}
B & -C\\
C & B
\end{array}
\right)
\left(
\begin{array}{c}
x\\y
\end{array}
\right).
\end{equation}

\noindent
From Eqs. (\ref{BC_Constraints}) it is easy to see that the matrix rotating this column is orthogonal. Moreover, its determinant is equal to 1 because the $U_5$ group manifold is connected. Therefore, this matrix belongs to the group $SO_{10}$.

From Eq. (\ref{U5_Subgroup}) we see that the properly normalized generators of $SO_{10}$ corresponding to the subgroup $SU_5$ can be written in the form

\begin{equation}
t_A\bigg|_{SU_5\subset SO_{10}} = \left\{
\begin{array}{l}
{\displaystyle \frac{1}{\sqrt{2}}\left(
\begin{array}{cc}
t_{A,5} & 0\\
0 & t_{A,5}
\end{array}
\right) = \frac{1}{\sqrt{2}}\, T(t_{A,5}),}\qquad\quad \text{if $t_{A,5}$ is purely imaginary};\\
\vphantom{1}\\
{\displaystyle \frac{i}{\sqrt{2}}\left(
\begin{array}{cc}
0 & t_{A,5}\\
- t_{A,5} & 0
\end{array}
\right)= \frac{1}{\sqrt{2}}\, T(t_{A,5}),}\qquad\, \text{if $t_{A,5}$ is real},
\end{array}
\right.
\end{equation}

\noindent
where $t_{A,5}$ (with $A=1,\ldots,24$) are the generators of the $SU_5$ fundamental representation normalized by the condition

\begin{equation}
\mbox{tr}\big(t_{A,5} t_{B,5}\big) = \frac{1}{2}\delta_{AB}.
\end{equation}

\noindent
(Certainly, it is convenient to choose them in such a way that they are either real or purely imaginary). The factor $1/\sqrt{2}$ appears due to the normalization condition (\ref{Metric_Definition}) in which the metric is equal to $\delta_{\bm{AB}}$ (for both $SO_{10}$ and $SU_5$). Due to this factor the coupling constants for the group $SO_{10}$ and its $SU_5$ subgroup are related by the equation

\begin{equation}\label{SU5_Coupling}
e_5 = \frac{e_{10}}{\sqrt{2}}.
\end{equation}

Similarly, from Eq. (\ref{U5_Subgroup}) we see that the generator of the $U_1$ subgroup (again normalized by the condition (\ref{Metric_Definition})) is

\begin{equation}\label{T_2^10}
t\Big|_{U_1\subset SO_{10}} = -\frac{i}{\sqrt{20}} \left(
\begin{array}{cc}
0 & -1_5\\
1_5 & 0
\end{array}
\right).
\end{equation}

According to \cite{Slansky:1981yr} the branching rules for the lowest $SO_{10}$ representations have the form

\begin{eqnarray}\label{SO10_Branching_Rules}
&& 10\Big|_{SO_{10}} = 5(2) + \,\xbar{5}\,(-2)\Big|_{SU_5\times U_1};\nonumber\\
&& 16\Big|_{SO_{10}} = 1(-5)+\,\xbar{5}\,(3) + 10(-1)\Big|_{SU_5\times U_1};\nonumber\\
&& \xbar{16}\Big|_{SO_{10}} = 1(5) + 5(-3) + \,\xbar{10}\,(1)\Big|_{SU_5\times U_1};\nonumber\\
&& 45\Big|_{SO_{10}} = 1(0) + 10(4) + \,\xbar{10}\,(-4) + 24(0)\Big|_{SU_5\times U_1},
\end{eqnarray}

\noindent
where the $U_1$ charges are (as usual) presented in the round brackets. It is easy to see that they can be identified with the eigenvalues of the operator (\ref{T_2^10}) multiplied by $4\sqrt{5}$.

\subsection{Options for symmetry breaking}
\hspace*{\parindent}\label{Subsection_SO10_To_SU5_Options}

Evidently, the group $SO_{10}\times U_1$ contains the subgroup $SU_5\times U_1\times U_1$ which is constructed as

\begin{equation}\label{SU5_U1^2_Subgroup}
SO_{10}\times \underbrace{U_1}_{\beta_1^{(10)}} \supset (SU_{5}\times \underbrace{U_1}_{\beta_2^{(10)}})\times \underbrace{U_1}_{\beta_1^{(10)}},
\end{equation}

\noindent
where $\beta_1^{(10)}$ and $\beta_2^{(10)}$ are real numbers parameterizing the transformations of the $U_1$ subgroups. According to the branching rule for the $SO_{10}$ representation $16$ presented in the second line of Eq. (\ref{SO10_Branching_Rules}), a vacuum expectation value of a scalar field in this representation can break the symmetry down to $SU_5 \times U_1$, where the $U_1$ subgroup of the little group is a nontrivial combination of the transformations of two $U_1$ factors in (\ref{SU5_U1^2_Subgroup}). The explicit form of this combination and the corresponding coupling constant essentially depend on two $U_1$ charges of the scalar field which acquires a vacuum expectation value. As we have already mentioned above, there are three different options for the $SO_{10}\times U_1\to SU_5\times U_1$ symmetry breaking which are separately considered below.

\bigskip

{\bf B-1-1. Vacuum expectation values of $27(-1)\Big|_{E_6\times U_1}$ and $16(-1)\Big|_{SO_{10}\times U_1}$.}

\medskip

In this case the vacuum expectation values are acquired by scalar fields in the representations $27(-1)$ of $E_6\times U_1$ and $16(-1)$ of $SO_{10}\times U_1$ (and/or the corresponding conjugated representations). To investigate the symmetry breaking, we start with the branching rule of the representation $16(-1)$ with respect to the subgroup (\ref{SU5_U1^2_Subgroup}),

\begin{equation}\label{16_B11_Branching_Rule}
16(-1)\Big|_{SO_{10}\times U_1} = 10(-1,-1) + \,\xbar{5}\,(-1,3) + \bm{1(-1,-5)}\Big|_{SU_{5}\times U_1\times U_1}.
\end{equation}

\noindent
From this equation we see that the original $SO_{10}\times U_1$ symmetry can be broken down to $SU_5\times U_1$, where the $U_1$ charge with respect to the little group is related to the charges with respect to the $U_1\times U_1$ subgroup in Eq. (\ref{SU5_U1^2_Subgroup}) as

\begin{equation}\label{SU5_B11_Little_Group_Charge}
Q_1^{(5)} = \frac{1}{4} \Big(5 Q_1^{(10)} - Q_2^{(10)} \Big).
\end{equation}

\noindent
Really, in this case the $SU_5$ singlet (indicated in Eq. (\ref{16_B11_Branching_Rule}) by the bold font) has the charge $\frac{1}{4}(5\cdot (-1)- (-5)) = 0$. The coefficient $1/4$ in this equation is introduced for the convenience of notation.

The branching rule for the $E_8$ representation $248$ with respect to the $SU_5\times U_1\times U_1$ subgroup (\ref{SU5_U1^2_Subgroup}) can be derived using Eqs. (\ref{248_To_SO10_B1}) and (\ref{SO10_Branching_Rules}),

\begin{eqnarray}\label{E8_SU5_B1_Preliminary}
&&\hspace*{-5mm} 248\Big|_{E_8} = 10\times 1(0,0) + 3\times 1(4,0) + 3\times 1(-4,0) + 3\times 1(1,5) + 3\times 1(-1,-5) +1(3,-5)\nonumber\\
&&\hspace*{-5mm} \qquad\quad\ + 1(-3,5) + 3\times 5(1,-3) + 3\times\,\xbar{5}\,(-1,3) + 3\times 5(-2,2) + 3\times \,\xbar{5}\,(2,-2) + 3\times 5(2,2) \vphantom{\Big(}\nonumber\\
&&\hspace*{-5mm} \qquad\quad\ + 3\times \,\xbar{5}\,(-2,-2) + 5(-3,-3) + \,\xbar{5}\,(3,3) +10(0,4) +\,\xbar{10}\,(0,-4) + 10(3,-1) + \,\xbar{10}\,(-3,1) \vphantom{\Big(}\nonumber\\
&&\hspace*{-5mm} \qquad\quad\ + 3\times 10(-1,-1) + 3\times \,\xbar{10}\,(1,1) + 24(0,0) \Big|_{SU_5\times U_1\times U_1}.\qquad
\end{eqnarray}

\noindent
The charges in the brackets $(Q_1^{(10)}, Q_2^{(10)})$ correspond to the $U_1$ groups parameterized by $\beta_1^{(10)}$ (the first number) and by $\beta_2^{(10)}$ (the second number). Taking into account that, according to Eq. (\ref{SO10_U1_B1_Coupling}),

\begin{equation}
e_1^{(10)} = \frac{e_{10}}{4\sqrt{3}},
\end{equation}

\noindent
we see that the first number ($Q_1^{(10)}$) is an eigenvalue of the operator $4\sqrt{3}\,t_1^{(10)}$, where $t_1^{(10)}$ is the generator of the $U_1$ factor in group $SO_{10}\times U_1$ normalized in the same way as the $SO_{10}$ generators. As we have already mentioned in the end of Sect. \ref{Subsection_SO10_To_SU5_Embedding}, the second number ($Q_2^{(10)}$) is an eigenvalue of the operator

\begin{equation}\label{Q_2^10_Operator}
4\sqrt{5}\, t\Big|_{U_1\subset SO_{10}}.
\end{equation}

\noindent
Therefore, the little group charge $Q_1^{(5)}$ is an eigenvalue of the operator

\begin{equation}
\frac{1}{4}\bigg(5\cdot 4\sqrt{3}\, t_1^{(10)} - 4\sqrt{5}\, t\Big|_{U_1\subset SO_{10}} \bigg) = 4\sqrt{5} \bigg(\frac{\sqrt{15}}{4}\, t_1^{(10)} - \frac{1}{4}\, t\Big|_{U_1\subset SO_{10}} \bigg).
\end{equation}

\noindent
Taking into account that the rightmost operator in parentheses is obviously normalized in the same way as the $SO_{10}$ generators we see that the coupling constant for the $U_1$ subgroup of the little group $SU_5\times U_1$ is given by the equation

\begin{equation}
e_1^{(5)} = \frac{e_{10}}{4\sqrt{5}} = \frac{e_5}{2\sqrt{10}},
\end{equation}

\noindent
where we also took into account Eq. (\ref{SU5_Coupling}). Therefore, for the considered pattern of symmetry breaking we obtain

\begin{equation}
\mbox{B-1-1}:\qquad \alpha_5 = \frac{\alpha_{10}}{2};\qquad \alpha_1^{(5)} = \frac{\alpha_5}{40}.
\end{equation}

\noindent
Calculating the little group charges for all terms in Eq. (\ref{E8_SU5_B1_Preliminary}) according to Eq. (\ref{SU5_B11_Little_Group_Charge}) we construct the branching rule for the $E_8$ representation $248$ with respect to the little group $SU_5\times U_1$,

\begin{eqnarray}\label{248_To_SU5_B11}
&&\hspace*{-5mm} 248\Big|_{E_8} = 16\times 1(0) + 4\times 1(5) + 4\times 1(-5) + 6\times 5(2) + 6\times \,\xbar{5}\,(-2) + 4\times 5(-3) \nonumber\\
&&\hspace*{-5mm} \qquad\quad\ + 4\times \,\xbar{5}\,(3) + {10}(4) + \,\xbar{10}\,(-4)
+ 4\times 10(-1) + 4\times \xbar{10}\,(1) + 24(0) \Big|_{SU_5\times U_1}.\qquad
\end{eqnarray}

Below we will see that the next symmetry breaking $SU_5\times U_1 \to SU_3\times SU_2\times U_1$ can be realized by a vacuum expectation value of a scalar field in the representation $10$ of $SU_5$. Therefore, from Eq. (\ref{248_To_SU5_B11}) we conclude that there are two different options for this symmetry breaking, namely, either by a vacuum expectation value of $10(-1)$ (and/or $\,\xbar{10}\,(1)$) or by a vacuum expectation value of $10(4)$ (and/or $\,\xbar{10}\,(-4)$). We will denote them B-1-1-1 and B-1-1-2, respectively, following the convention that an option with the minimal absolute value of the $U_1$ charge is indicated by the number 1.

\bigskip

{\bf B-1-2. Vacuum expectation values of $27(-1)\Big|_{E_6\times U_1}$ and $16(3)\Big|_{SO_{10}\times U_1}$.}

\medskip

Another option for the symmetry breaking appears if vacuum expectation values are acquired by scalar fields in the representations $27(-1)$ of $E_6\times U_1$ and $16(3)$ of $SO_{10}\times U_1$. The branching rule of the representation $16(3)$ with respect to the subgroup (\ref{SU5_U1^2_Subgroup}) reads as

\begin{equation}\label{16_B12_Branching_Rule}
16(3)\Big|_{SO_{10}\times U_1} = 10(3,-1) + \,\xbar{5}\,(3,3) + \bm{1(3,-5)}\Big|_{SU_{5}\times U_1\times U_1},
\end{equation}

\noindent
so that in this case the little group charge can be defined as

\begin{equation}\label{SU5_B12_Little_Group_Charge}
Q_1^{(5)} = \frac{1}{4} \Big(5 Q_1^{(10)} + 3Q_2^{(10)} \Big),
\end{equation}

\noindent
where the coefficient $1/4$ is included for convenience of notation. Really, in this case the charge of the singlet in Eq. (\ref{16_B12_Branching_Rule}) is $(5\cdot 3+ 3\cdot(-5))/4 = 0$, so that the representation $16(3)$ really contains a part invariant under the little group.

Similar to the case B-1-1, the charge (\ref{SU5_B12_Little_Group_Charge}) can be considered as an eigenvalue of the operator

\begin{equation}
\frac{1}{4}\bigg(5\cdot 4\sqrt{3}\, t_1^{(10)} + 3\cdot 4\sqrt{5}\, t\Big|_{U_1\subset SO_{10}} \bigg) = 2\sqrt{30} \bigg(\sqrt{\frac{5}{8}}\, t_1^{(10)} + \sqrt{\frac{3}{8}}\, t\Big|_{U_1\subset SO_{10}} \bigg).
\end{equation}

\noindent
Again the rightmost operator in parentheses is normalized in the same way as the $SO_{10}$ generators, so that the value of the coupling constant for the $U_1$ factor in the group $SU_5\times U_1$ can be related to the coupling constant $e_{10}$ by the equation

\begin{equation}
e_1^{(5)} = \frac{e_{10}}{2\sqrt{30}} = \frac{e_5}{\sqrt{60}}.
\end{equation}

\noindent
Therefore, for the considered symmetry breaking pattern the values of couplings are given by the expressions

\begin{equation}
\mbox{B-1-2}:\qquad \alpha_5 = \frac{\alpha_{10}}{2};\qquad \alpha_1^{(5)} = \frac{\alpha_5}{60}.
\end{equation}

Calculating the charge (\ref{SU5_B12_Little_Group_Charge}) for all terms in the expression (\ref{E8_SU5_B1_Preliminary}) we obtain the branching rule for the $E_8$ representation $248$ with respect to the little group $SU_5\times U_1$,

\begin{eqnarray}\label{248_To_SU5_B12}
&&\hspace*{-3mm} 248\Big|_{E_8} = 12\times 1(0) + 6\times 1(5) + 6\times 1(-5) + 6\times 5(-1) + 6\times \,\xbar{5}\,(1) + 3\times 5(4) + 3\times \,\xbar{5}\,(-4)\nonumber\\
&&\hspace*{-3mm} \qquad\quad\ +5(-6) + \,\xbar 5\,(6) + 3\times {10}(-2) + 3\times \,\xbar{10}\,(2) + 2\times 10(3) + 2\times \,\xbar{10}\,(-3) + 24(0)\Big|_{SU_5\times U_1}.\nonumber\\
\end{eqnarray}

From this expression we see that the next symmetry breaking can also be realized in two different ways, namely by vacuum expectation values of the representations $10(-2)$ (and/or $\,\xbar{10}\,(2)$) or $10(3)$ (and/or $\,\xbar{10}\,(-3)$). Following our usual conventions, we denote them as B-1-2-1 and B-1-2-2, respectively. Details of the corresponding symmetry breakings will be discussed below in Sect. \ref{Section_SU5_TO_MSSM}.

\bigskip

{\bf B-2-1. Vacuum expectation values of $27(2)\Big|_{E_6\times U_1}$ and $16(1)\Big|_{SO_{10}\times U_1}$.}

\medskip

The symmetry breaking $E_6\times U_1 \to SO_{10}\times U_1 \to SU_5\times U_1$ can also be realized by vacuum expectation values of scalar fields in the representations $27(2)$ of $E_6\times U_1$ and $16(1)$ of $SO_{10}\times U_1$. Due to the branching rule

\begin{equation}\label{16_B21_Branching_Rule}
16(1)\Big|_{SO_{10}\times U_1} = 10(1,-1) + \,\xbar{5}\,(1,3) + \bm{1(1,-5)}\Big|_{SU_{5}\times U_1\times U_1},
\end{equation}

\noindent
the little group charge can be chosen in the form

\begin{equation}\label{SU5_B21_Little_Group_Charge}
Q_1^{(5)} = -\frac{1}{2} \Big(5 Q_1^{(10)} + Q_2^{(10)} \Big),
\end{equation}

\noindent
where the factor $-1/2$ is introduced for the further convenience. Taking into account that for the B-2 pattern of $E_6\times U_1$ breaking

\begin{equation}
e_1^{(10)} = \frac{e_{10}}{4}
\end{equation}

\noindent
we see that $Q_1^{(10)}$ is an eigenvalue of the operator $4 t_1^{(10)}$, while $Q_2^{(10)}$ is again an eigenvalue of the operator (\ref{Q_2^10_Operator}). Therefore, the charge (\ref{SU5_B21_Little_Group_Charge}) is an eigenvalue of the operator

\begin{equation}
- \frac{1}{2}\bigg(5\cdot 4 t_1^{(10)} + 4\sqrt{5}\, t\Big|_{U_1\subset SO_{10}} \bigg) = 2\sqrt{30} \bigg(-\sqrt{\frac{5}{6}}\, t_1^{(10)} - \sqrt{\frac{1}{6}}\, t\Big|_{U_1\subset SO_{10}} \bigg).
\end{equation}

\noindent
This implies that the coupling constant for the $U_1$ component of the little group is related to the couplings $e_{10}$ and $e_5$ by the equation

\begin{equation}
e_1^{(5)} = \frac{e_{10}}{2\sqrt{30}} = \frac{e_5}{\sqrt{60}},
\end{equation}

\noindent
so that for the symmetry breaking pattern B-2-1 we obtain

\begin{equation}
\mbox{B-2-1}:\qquad \alpha_5 = \frac{\alpha_{10}}{2};\qquad \alpha_1^{(5)} = \frac{\alpha_5}{60}.
\end{equation}

\noindent
Note that these values of the coupling constants exactly coincide with the ones for the option B-1-2.

Also in this case it is necessary to derive the branching rule for the $E_8$ representation $248$ with respect to the little group $SU_5\times U_1$. For this purpose we first construct the branching rule with respect to the $SU_5\times U_1\times U_1$ subgroup (\ref{SU5_U1^2_Subgroup}) using Eqs. (\ref{248_To_SO10_B2}) and (\ref{SO10_Branching_Rules}),

\begin{eqnarray}
&&\hspace*{-5mm} 248\Big|_{E_8} = 8\times 1(0,0) + 4\times 1(2,0) + 4\times 1(-2,0) + 2\times 1(1,5) + 2\times 1(-1,-5) + 2\times 1(1,-5)\nonumber\\
&&\hspace*{-5mm} \qquad\quad\ + 2\times 1(-1,5) + 2\times 5(1,-3) + 2\times \,\xbar{5}\,(-1,3) + 2\times 5(-1,-3) + 2\times \,\xbar{5}\,(1,3) + 5(2,2) \vphantom{\Big(}\nonumber\\
&&\hspace*{-5mm} \qquad\quad\ + \,\xbar{5}\,(-2,-2) + 5(-2,2) + \,\xbar{5}\,(2,-2) + 4\times 5(0,2)  + 4\times \,\xbar{5}\,(0,-2)  + 10(0,4) + \,\xbar{10}\,(0,-4) \vphantom{\Big(}\nonumber\\
&&\hspace*{-5mm} \qquad\quad\ + 2\times 10(-1,-1) + 2\times \,\xbar{10}\,(1,1) + 2\times 10(1,-1) + 2\times\, \xbar{10}\,(-1,1)
+ 24(0,0)\Big|_{SU_5\times U_1\times U_1}.\nonumber\\
\end{eqnarray}

\noindent
After that, for each term we calculate the charge (\ref{SU5_B21_Little_Group_Charge}) corresponding to the $U_1$ subgroup of the little group. The result is given by the expression

\begin{eqnarray}\label{248_To_SU5_B21}
&&\hspace*{-3mm} 248\Big|_{E_8} = 12\times 1(0) + 6\times 1(5) + 6\times 1(-5) + 6\times 5(-1) + 6\times \,\xbar{5}\,(1) + 3\times 5(4) + 3\times \,\xbar{5}\,(-4)\nonumber\\
&&\hspace*{-3mm} \qquad\quad\ +5(-6) + \,\xbar 5\,(6) + 3\times {10}(-2) + 3\times \,\xbar{10}\,(2) + 2\times 10(3) + 2\times \,\xbar{10}\,(-3) + 24(0)\Big|_{SU_5\times U_1}.\nonumber\\
\end{eqnarray}

\noindent
As well as the values of the coupling constants, this branching rule also coincides with the one for the case B-1-2. Therefore, the further symmetry breaking ($SU_5\times U_1 \to SU_3\times SU_2\times U_1$) occurs in the same way as for B-1-2. That is why below in Sect. \ref{Section_SU5_TO_MSSM} we will not consider it separately and present only the results.

\section{The symmetry breaking $SU_5\times U_1 \to SU_3\times SU_2\times U_1$}
\label{Section_SU5_TO_MSSM}

\subsection{The embedding $SU_3\times SU_2 \times U_1 \subset SU_5$}
\hspace*{\parindent}

The group $SU_3\times SU_2\times U_1$ can be embedded into $SU_5$ as

\begin{equation}
\omega_5 = \left(
\begin{array}{cc}
e^{-2i\beta_2^{(5)}} \omega_3 & 0 \\
0 &  e^{3i\beta_2^{(5)}} \omega_2
\end{array}
\right),
\end{equation}

\noindent
where $\omega_3\in SU_3$, $\omega_2\in SU_2$, and the real number $\beta_2^{(5)}$ is a parameter of the $U_1$ subgroup. If the $U_1$ charge is identified with the coefficient at $i\beta_2^{(5)}$ in the arguments of exponents, then the branching rules for some lowest $SU_5$ representation are written as

\begin{eqnarray}\label{SU5_Branching_Rules}
&& 5\Big|_{SU_5} = [1,2](3) + [3,1](-2)\Big|_{SU_3\times SU_2\times U_1};\nonumber\\
&&\, \xbar{5}\,\Big|_{SU_5} = [1,2](-3) + [\,\xbar{3}\,,1](2)\Big|_{SU_3\times SU_2\times U_1};\nonumber\\
&& 10\Big|_{SU_5} = [1,1](6) + [\,\xbar{3}\,,1](-4) + [3,2](1)\Big|_{SU_3\times SU_2\times U_1};\nonumber\\
&&\, \xbar{10}\,\Big|_{SU_5} = [1,1](-6) + [3,1](4) + [\,\xbar{3}\,,2](-1)\Big|_{SU_3\times SU_2\times U_1};\nonumber\\
&& 24\Big|_{SU_5} = [1,1](0) + [1,3](0) + [3,2](-5) + [\,\xbar{3}\,,2](5) + [8,1](0)\Big|_{SU_3\times SU_2\times U_1},\qquad
\end{eqnarray}

\noindent
where the first symbol in the square brackets denotes the $SU_3$ representation, the second one corresponds to the $SU_2$ representation, and the $U_1$ charge is presented in the round brackets.

The (properly normalized) $SU_5$ generators corresponding to the $SU_3$, $SU_2$, and $U_1$ subgroups can be presented as

\begin{eqnarray}\label{SU3_SU2_U1_Generators}
&& t_{A}\Big|_{SU_3\subset SU_5} = \frac{1}{2}\left(
\begin{array}{cc}
\lambda_A & 0\\
0 & 0
\end{array}
\right);\qquad t_\alpha\Big|_{SU_2\subset SU_5} = \frac{1}{2}\left(
\begin{array}{cc}
0 & 0\\
0 & \sigma_\alpha
\end{array}
\right);\qquad\nonumber\\
&& \vphantom{1}\nonumber\\
&&\qquad\qquad\quad\ \ t\Big|_{U_1\subset SU_5} = \frac{1}{2\sqrt{15}} \left(
\begin{array}{cc}
-2\cdot 1_3 & 0\\
0 & 3\cdot 1_2
\end{array}
\right),
\end{eqnarray}

\noindent
where $\lambda_A$ with $A=1,\ldots,8$ and $\sigma_\alpha$ with $\alpha=1,2,3$ are the Gell-Mann and Pauli matrices, respectively. This in particular implies that $e_5=e_3=e_2$. From the form of the $U_1$ generator we also see that the numbers presented in Eq. (\ref{SU5_Branching_Rules}) in the round brackets are the eigenvalues of the operator $\sqrt{60}\, t\Big|_{U_1\subset SU_5}$ (in a relevant representation).

Thus, for the simple subgroups the coupling constants are related by the equations

\begin{equation}
e_2 = e_3 = e_5 = \frac{e_{10}}{\sqrt{2}} = \frac{e_6}{\sqrt{6}} = \frac{e_7}{\sqrt{12}} = \frac{e_8}{\sqrt{60}}
\end{equation}

\noindent
or, equivalently, (for $\alpha_i\equiv e_i^2/4\pi$)

\begin{equation}
\alpha_2 = \alpha_3 = \alpha_5 = \frac{\alpha_{10}}{2} = \frac{\alpha_6}{6} = \frac{\alpha_7}{12} = \frac{\alpha_8}{60}.
\end{equation}

\subsection{Options for symmetry breaking}
\hspace*{\parindent}

{\bf B-1-1-1. Vacuum expectation values of $27(-1)\Big|_{E_6\times U_1}$, $16(-1)\Big|_{SO_{10}\times U_1}$, and $10(-1)\Big|_{SU_5\times U_1}$.}

\medskip

Obviously, the group $SU_5\times U_1$ contains the subgroup $SU_3\times SU_2\times U_1\times U_1$,

\begin{equation}\label{SU3_SU2_U1^2_Subgroup}
SU_{5}\times \underbrace{U_1}_{\beta_1^{(5)}} \supset (SU_{3}\times SU_2 \times \underbrace{U_1}_{\beta_2^{(5)}})\times \underbrace{U_1}_{\beta_1^{(5)}}.
\end{equation}

\noindent
We are interested in the corresponding branching rule for the representation $10(-1)$ because this representation is present in Eq.  (\ref{248_To_SU5_B11}) and can be used for the further symmetry breaking. From Eq. (\ref{SU5_Branching_Rules}) we obtain

\begin{equation}\label{10_B111_Branching_Rule}
10(-1)\Big|_{SU_5\times U_1} = \bm{[1,1](-1,6)} + [\,\xbar{3},1](-1,-4) + [3,2](-1,1)\Big|_{SU_3\times SU_2\times U_1\times U_1},
\end{equation}

\noindent
where the charge $Q_1^{(5)}$ corresponding to the $U_1$ group with the parameter $\beta_1^{(5)}$ is given by the first number in the round brackets, while the second number is the charge $Q_2^{(5)}$ with respect to the $U_1$ group parameterized by $\beta_2^{(5)}$. We see that the right hand side of Eq. (\ref{10_B111_Branching_Rule}) contains an $SU_3\times SU_2$ singlet, which is indicated by the bold font. Looking at its charges with respect to $U_1\times U_1$ we conclude that the symmetry is broken down to $SU_3\times SU_2\times U_1$, where the $U_1$ charge of the little group can be chosen in the form

\begin{equation}\label{SU3_SU2_B111_Little_Group_Charge}
Q_1^{(Y)} = - \frac{1}{30} \Big(6 Q_1^{(5)} + Q_2^{(5)} \Big).
\end{equation}

\noindent
The factor $-1/30$ is again included for the convenience of notation. Earlier we chose similar coefficients in such a way that the little group charges will be the least possible integers. In contrast, now we take such a normalization factor that the $U_1$ charges of various superfields can be identified with the hypercharge in MSSM, see Eq. (\ref{248_To_SU3_SU2_B111}) below.

In Sect. \ref{Subsection_SO10_To_SU5_Options} we demonstrated that for the option B-1-1 the coupling constant for the $U_1$ factor in $SU_5\times U_1$ is

\begin{equation}
e_1^{(5)} = \frac{e_{5}}{2\sqrt{10}}.
\end{equation}

\noindent
This implies that the charge $Q_1^{(5)}$ is an eigenvalue of the operator $2\sqrt{10}\, t_1^{(5)}$, where $t_1^{(5)}$ is a generator of the $U_1$ factor in $SU_5\times U_1$ normalized in the same way as the generators of the $SU_5$ factor. As we have already mentioned, the charge $Q_2^{(5)}$ can be considered as an eigenvalue of the operator

\begin{equation}\label{Q_2^5_Operator}
2\sqrt{15}\, t\Big|_{U_1\subset SU_5},
\end{equation}

\noindent
where $t\Big|_{U_1\subset SU_5}$ is defined by the last equation in (\ref{SU3_SU2_U1_Generators}). Therefore, the charge (\ref{SU3_SU2_B111_Little_Group_Charge}) can be viewed as an eigenvalue of the operator

\begin{equation}
- \frac{1}{30} \bigg(6\cdot 2\sqrt{10}\, t_1^{(5)} + 2\sqrt{15}\, t\Big|_{U_1\subset SU_5} \bigg) = \sqrt{\frac{5}{3}} \bigg(-\frac{2\sqrt{6}}{5}\, t_1^{(5)} - \frac{1}{5}\, t\Big|_{U_1\subset SU_5} \bigg).
\end{equation}

\noindent
Taking into account that the rightmost operator in the round brackets is normalized in the same way as the $SU_5$ generators we obtain that the coupling constant for the $U_1$ factor in the little group $SU_3\times SU_2\times U_1$ is

\begin{equation}
e_1^{(Y)} = \sqrt{\frac{3}{5}}\, e_5
\end{equation}

\noindent
or, equivalently,

\begin{equation}
\mbox{B-1-1-1:}\qquad \alpha_2 = \alpha_3 = \alpha_5;\qquad \alpha_1^{(Y)} = \frac{3}{5} \alpha_5,
\end{equation}

\noindent
where $\alpha_1^{(Y)}\equiv \big(e_1^{(Y)}\big)^2/4\pi$. This implies that we obtain the standard coupling constant unification $\alpha_1=\alpha_2=\alpha_3$ with $\alpha_1\equiv 5\alpha_1^{(Y)}/3$, so that in this case

\begin{equation}
\sin^2\theta_W=\frac{3}{8}.
\end{equation}

\noindent
Below we will see that B-1-1-1 is the only option which allows achieving these relations for the coupling constants. Various details of the symmetry breaking for this case are presented in Table \ref{Table_Summary_B-1-1-1}.

\begin{table}[h]
\begin{center}
\begin{tabular}{|c|c|c|c|c|c|c|c|}
\hline
Group $\vphantom{\Big(}$ & \multicolumn{6}{|c|}{\bf{Quantum numbers}} & \bf{Couplings} \\
\hline
$E_8$ $\vphantom{\Big(}$ & \multicolumn{6}{|c|}{$\bm{248}$} & $e_8$ \\
\hline
$E_7\times U_1$ $\vphantom{\Big(}$ & $1(0)$ & $1(2)$ & $1(-2)$ & $133(0)$ & $\bm{56(1)}$ & $56(-1)$ & $e_7=e_8/\sqrt{5}$ \\
$\vphantom{\Big(}$                 &        &        &         &          &                &            & $\big(e_1^{(7)} = e_7/4\sqrt{3}\big)$ \\
\hline
$E_6\times U_1$ $\vphantom{\Big(}$ & $1(0)$ & $1(-3)$ & $1(3)$ & $78(0)$                                    & $\hspace*{0.5mm}\bm{27(-1)}\hspace*{0.5mm}$          & $\,\xbar{27}\,(1)$                          & $e_6=e_7/\sqrt{2}$            \\
$\vphantom{\Big(}$                 &        &        &         & $1(0)$                                     & $\,\xbar{27}\,(-2)$ & $27(2)$  & $\big(e_1^{(6)}=e_6/6\sqrt{2}\big)$ \\
$\vphantom{\Big(}$                 &        &        &         & $\hspace*{0.5mm}\bm{27(-1)}\hspace*{0.5mm}$  & $1(0)$           & $1(0)$                                    &                               \\
$\vphantom{\Big(}$                 &        &        &         & $\,\xbar{27}\,(1)$                           & $1(-3)$           & $1(3)$                                     &                               \\
\hline
$SO_{10}\times U_1$ $\vphantom{\Big(}$ & \multicolumn{6}{|l|}{$9\times 1(0) + 3\times 1(4) + 3\times 1(-4) + 45(0) +3\times 10(-2)$}
& $e_{10} = e_6/\sqrt{3}$\\
$\vphantom{\Big(}$ & \multicolumn{6}{|l|}{$ + 3\times 10(2) + 3\times \bm{16(-1)} + 3\times \,\xbar{16}\,(1) + 16(3) + \,\xbar{16}\,(-3) $} & $\big(e_1^{(10)} = e_{10}/4\sqrt{3}\big)$\\
\hline
$SU_{5}\times U_1$ $\vphantom{\Big(}$ & \multicolumn{6}{|l|}{$16\times 1(0) + 4\times 1(5) + 4\times 1(-5) + 24(0) + 6\times 5(2) $} & $e_{5} = e_{10}/\sqrt{2}$\\
$\vphantom{\Big(}$ & \multicolumn{6}{|l|}{$ + 6\times \xbar{5}\,(-2) + 4\times 5(-3) + 4\times \xbar{5}\,(3) + {10}(4) + \xbar{10}\,(-4)$} & $\big(e_1^{(5)}=e_{5}/2\sqrt{10}\big)$\\
$\vphantom{\Big(}$ & \multicolumn{6}{|l|}{$ + 4\times \bm{10(-1)} + 4\times \xbar{10}\,(1) $} & \\
\hline
\hspace*{-1mm}$SU_3\times SU_2\times U_1$\hspace*{-1mm} & \multicolumn{6}{|l|}{$25\times\bm{[1,1](0)} + 5\times[1,1](1) +5\times\bm{[1,1](-1)} $} & $\bm{e_3 = e_2 =e_5}\vphantom{\Big(}$ \\
$\vphantom{\Big(}$ & \multicolumn{6}{|l|}{$ + [1,3](0) + 10\times\bm{[1,2](1/2)} + 10\times\bm{[1,2](-1/2)} $} & \hspace*{-1mm}$\bm{\big(e_1^{(Y)} = e_{5} \sqrt{3/5}\big)}$\hspace*{-1mm}\\
$\vphantom{\Big(}$ & \multicolumn{6}{|l|}{$ +10\times \bm{[3,1](-1/3)} + 10\times[\,\xbar{3}\,,1](1/3) + 5\times \bm{[3,1](2/3)}  $} & $\bm{\text{\bf{sin}}^2\theta_W=3/8}$ \\
$\vphantom{\Big(}$ & \multicolumn{6}{|l|}{$ + 5\times [\,\xbar{3}\,,1](-2/3) + 5\times [3,2](1/6) + 5\times \bm{[\,\xbar{3}\,,2](-1/6)} $} & \\
$\vphantom{\Big(}$ & \multicolumn{6}{|l|}{$ +[3,2](-5/6) +[\,\xbar{3}\,,2](5/6)  + [8,1](0)$} & \\
\hline
\end{tabular}
\end{center}
\caption{The option {\bf{B-1-1-1}} for the decomposition of 248 into irreducible representations of various subgroups appearing in the symmetry breaking chain. This is the only option for which the unification of couplings occurs according to Eq. (\ref{Standard_Coupling_Unification}) and the final branching rule includes all representations needed for accommodating the MSSM superfields. Note that the absolute values of $U_1$ charges for all (super)fields acquiring vacuum expectation values in this case are the minimal possible.}
\label{Table_Summary_B-1-1-1}
\end{table}

To construct the branching rule for the $E_8$ representation $248$ with respect to the little group $SU_3\times SU_2\times U_1$, we start with the branching rule with respect to the $SU_3\times SU_2\times U_1\times U_1$ subgroup (\ref{SU3_SU2_U1^2_Subgroup}). This branching rule can be derived using Eqs. (\ref{248_To_SU5_B11}) and (\ref{SU5_Branching_Rules}),

\begin{eqnarray}\label{E8_SU3_SU2_B11_Preliminary}
&&\hspace*{-5mm} 248\Big|_{E_8} = 17\times [1,1](0,0) + [1,1](4,6) + [1,1](-4,-6) + 4\times [1,1](5,0) + 4\times [1,1](-5,0)\nonumber\\
&&\hspace*{-5mm}\qquad\quad\ + 4\times [1,1](-1,6) + 4\times [1,1](1,-6) + 6\times [1,2](2,3) + 6\times [1,2](-2,-3)\vphantom{\Big(}\nonumber\\
&&\hspace*{-5mm}\qquad\quad\ + 4\times [1,2](-3,3) + 4\times [1,2](3,-3) + [1,3](0,0) + [3,1](-4,4) + [\,\xbar{3}\,,1](4,-4)\vphantom{\Big(}\nonumber\\
&&\hspace*{-5mm}\qquad\quad\ + 4\times [3,1](-3,-2) + 4\times [\,\xbar{3}\,,1](3,2) + 6\times [3,1](2,-2) + 6\times [\,\xbar{3}\,,1](-2,2) \vphantom{\Big(}\nonumber\\
&&\hspace*{-5mm}\qquad\quad\ + 4\times [3,1](1,4) + 4\times [\,\xbar{3}\,,1](-1,-4) + [3,2](0,-5) + [\,\xbar{3}\,,2](0,5) + 4\times [3,2](-1,1)\vphantom{\Big(}\nonumber\\
&&\hspace*{-5mm}\qquad\quad\ + 4\times [\,\xbar{3}\,,2](1,-1) + [3,2](4,1) + [\,\xbar{3}\,,2](-4,-1) + [8,1](0,0)\Big|_{SU_3\times SU_2\times U_1\times U_1}.
\end{eqnarray}

\noindent
Next, for all terms we calculate the charges with respect to the $U_1$ subgroup of the little group with the help of Eq. (\ref{SU3_SU2_B111_Little_Group_Charge}). The result is written as

\begin{eqnarray}\label{248_To_SU3_SU2_B111}
&&\hspace*{-5mm} 248\Big|_{E_8} = 25\times\bm{[1,1](0)} + 5\times[1,1](1) +5\times\bm{[1,1](-1)} + [1,3](0) + 10\times\bm{[1,2](1/2)}\nonumber\\
&&\hspace*{-5mm} \qquad\quad\ + 10\times\bm{[1,2](-1/2)} +10\times \bm{[3,1](-1/3)} + 10\times[\,\xbar{3}\,,1](1/3) + 5\times \bm{[3,1](2/3)}\vphantom{\Big(}\nonumber\\
&&\hspace*{-5mm} \qquad\quad\ + 5\times [\,\xbar{3}\,,1](-2/3) + 5\times [3,2](1/6) + 5\times \bm{[\,\xbar{3}\,,2](-1/6)} +[3,2](-5/6) +[\,\xbar{3}\,,2](5/6)\vphantom{\Big(}\nonumber\\
&&\hspace*{-5mm} \qquad\quad\ + [8,1](0) \Big|_{SU_3\times SU_2\times U_1},
\end{eqnarray}

\noindent
where the parts corresponding to the MSSM chiral superfields (including Higgses and right neutrinos) are indicated by the bold font. From this equation we see that the right hand side contains terms with the quantum numbers of all MSSM superfields (including the right neutrinos), and the $U_1$ charge really coincides with the hypercharge due to the proper choice of the normalization factor in Eq. (\ref{SU3_SU2_B111_Little_Group_Charge}). Certainly, in this paper we do not discuss which fields survive after various symmetry breakings and how much representations $248$ we need for constructing MSSM (if possible). Here we only demonstrate that there is such an option for the symmetry breaking that the coupling constants of the resulting $SU_3\times SU_2\times U_1$ theory satisfy the standard unification conditions (\ref{Standard_Coupling_Unification}). Certainly, a more accurate analysis requires investigation of quantum corrections, which essentially depend on the scales of various symmetry breakings due to the threshold effects. Although these effects change the running of couplings, we can assume that all scales appearing in the symmetry breaking pattern under consideration are rather large and close to each other, so that the unification of couplings in the supersymmetric case persists at the quantum level. Nevertheless, this issue deserves a detailed separate study.

\bigskip

{\bf B-1-1-2. Vacuum expectation values of $27(-1)\Big|_{E_6\times U_1}$, $16(-1)\Big|_{SO_{10}\times U_1}$, and $10(4)\Big|_{SU_5\times U_1}$.}

\medskip

Alternatively, the symmetry breaking $SU_5\times U_1 \to SU_3\times SU_2\times U_1$ can be realized if the vacuum expectation value is acquired by the representation $10(4)$ (and/or $\,\xbar{10}\,(-4)$) present in Eq. (\ref{248_To_SU5_B11}). Its branching rule with respect to the $SU_3\times SU_2\times U_1\times U_1$ subgroup (\ref{SU3_SU2_U1^2_Subgroup}) reads as

\begin{equation}\label{10_B112_Branching_Rule}
10(4)\Big|_{SU_5\times U_1} = \bm{[1,1](4,6)} + [\,\xbar{3},1](4,-4) + [3,2](4,1)\Big|_{SU_3\times SU_2\times U_1\times U_1}.
\end{equation}

\noindent
From this equation we conclude that the charge with respect to the $U_1$ component of the little group $SU_3\times SU_2\times U_1$ can be written as

\begin{equation}\label{SU3_SU2_B112_Little_Group_Charge}
Q_1^{(Y)} = \frac{1}{30} \Big(- 3 Q_1^{(5)} + 2 Q_2^{(5)} \Big),
\end{equation}

\noindent
where we again choose the normalization factor in such a way that the resulting decomposition of the representation $248$ contains terms with the same values of the hypercharge as the MSSM superfields. The charge (\ref{SU3_SU2_B112_Little_Group_Charge}) is obtained as an eigenvalue of the operator

\begin{equation}\label{SU3_SU2_Operator}
\frac{1}{30} \bigg(- 3\cdot 2\sqrt{10}\, t_1^{(5)} +2 \cdot 2\sqrt{15}\, t\Big|_{U_1\subset SU_5} \bigg) = \sqrt{\frac{2}{3}} \bigg(-\sqrt{\frac{3}{5}}\, t_1^{(5)} + \sqrt{\frac{2}{5}}\, t\Big|_{U_1\subset SU_5} \bigg),
\end{equation}

\begin{table}[h]
\begin{center}
\begin{tabular}{|c|c|c|c|c|c|c|c|}
\hline
Group $\vphantom{\Big(}$ & \multicolumn{6}{|c|}{Quantum numbers} & Couplings \\
\hline
$E_8$ $\vphantom{\Big(}$ & \multicolumn{6}{|c|}{$\bm{248}$} & $e_8$ \\
\hline
$E_7\times U_1$ $\vphantom{\Big(}$ & $1(0)$ & $1(2)$ & $1(-2)$ & $133(0)$ & $\bm{56(1)}$ & $56(-1)$ & $e_7=e_8/\sqrt{5}$ \\
$\vphantom{\Big(}$                 &        &        &         &          &                &            & $\big(e_1^{(7)} = e_7/4\sqrt{3}\big)$ \\
\hline
$E_6\times U_1$ $\vphantom{\Big(}$ & $1(0)$ & $1(-3)$ & $1(3)$ & $78(0)$          & $\hspace*{0.5mm}\bm{27(-1)}\hspace*{0.5mm}$        & $\,\xbar{27}\,(1)$ & $e_6=e_7/\sqrt{2}$            \\
$\vphantom{\Big(}$                 &        &        &         & $1(0)$           & $\,\xbar{27}\,(-2)$  & $27(2)$  & $\big(e_1^{(6)}=e_6/6\sqrt{2}\big)$ \\
$\vphantom{\Big(}$                 &        &        &         & $\hspace*{0.5mm}\bm{27(-1)}\hspace*{0.5mm}$  & $1(0)$           & $1(0)$          &                               \\
$\vphantom{\Big(}$                 &        &        &         & $\,\xbar{27}\,(1)$  & $1(-3)$           & $1(3)$           &                               \\
\hline
$SO_{10}\times U_1$ $\vphantom{\Big(}$ & \multicolumn{6}{|l|}{$9\times 1(0) + 3\times 1(4) + 3\times 1(-4) + 45(0) +3\times 10(-2)$}
& $e_{10} = e_6/\sqrt{3}$\\
$\vphantom{\Big(}$ & \multicolumn{6}{|l|}{$ + 3\times 10(2) + 3\times \bm{16(-1)} + 3\times \,\xbar{16}\,(1) + 16(3) + \,\xbar{16}\,(-3) $} & $\big(e_1^{(10)} = e_{10}/4\sqrt{3}\big)$\\
\hline
$SU_{5}\times U_1$ $\vphantom{\Big(}$ & \multicolumn{6}{|l|}{$16\times 1(0) + 4\times 1(5) + 4\times 1(-5) + 24(0) + 6\times 5(2) $} & $e_{5} = e_{10}/\sqrt{2}$\\
$\vphantom{\Big(}$ & \multicolumn{6}{|l|}{$ + 6\times \,\xbar{5}\,(-2) + 4\times 5(-3) + 4\times \,\xbar{5}\,(3) + \bm{{10}(4)} + \,\xbar{10}\,(-4)$} & $\big(e_1^{(5)}=e_{5}/2\sqrt{10}\big)$\\
$\vphantom{\Big(}$ & \multicolumn{6}{|l|}{$ + 4\times 10(-1) + 4\times \,\xbar{10}\,(1) $} & \\
\hline
\hspace*{-1mm}$SU_3\times SU_2\times U_1$\hspace*{-1mm} & \multicolumn{6}{|l|}{$19\times\bm{[1,1](0)} + 8\times[1,1](1/2) +8\times[1,1](-1/2)$} & $e_3 = e_2 =e_5\vphantom{\Big(}$ \\
$\vphantom{\Big(}$ & \multicolumn{6}{|l|}{$ + 12\times [1,2](0)+ 4\times\bm{[1,2](1/2)} + 4\times\bm{[1,2](-1/2)} $} & \hspace*{-1mm}$\big(e_1^{(Y)} = e_{5}\sqrt{3/2}\big)$\hspace*{-1mm}\\
$\vphantom{\Big(}$ & \multicolumn{6}{|l|}{$ + [1,3](0) +8\times [3,1](1/6) + 8\times[\,\xbar{3}\,,1](-1/6) $} & $\sin^2\theta_W=3/5$ \\
$\vphantom{\Big(}$ & \multicolumn{6}{|l|}{$ + 6\times \bm{[3,1](-1/3)} + 6\times [\,\xbar{3}\,,1](1/3) + \bm{[3,1](2/3)}  $} & \\
$\vphantom{\Big(}$ & \multicolumn{6}{|l|}{$ + [\,\xbar{3}\,,1](-2/3) + 2\times [3,2](-1/3) + 2\times [\,\xbar{3}\,,2](1/3)  $} & \\
$\vphantom{\Big(}$ & \multicolumn{6}{|l|}{$ +4\times [3,2](1/6) + 4\times \bm{[\,\xbar{3}\,,2](-1/6)} + [8,1](0)$} & \\
\hline
\end{tabular}
\end{center}
\caption{Breaking of 248 into irreducible representations of various subgroups and the corresponding coupling constants for the option {\bf{B-1-1-2}}. In this case the coupling constants of the resulting $SU_3\times SU_2\times U_1$ theory do not satisfy the standard unification condition (\ref{Standard_Coupling_Unification}), and the final branching rule does not include the representation $[1,1](-1)$ needed for the right charged leptons.}\label{Table_Summary_B-1-1-2}
\end{table}

\noindent
where in the brackets in the right hand side we again extract the operator normalized in the same way as the generators of $SU_5$. From Eq. (\ref{SU3_SU2_Operator}) we conclude that the coupling constant for the $U_1$ component of the little group is related to the $SU_5$ coupling constant by the equation

\begin{equation}
e_1^{(Y)} = \sqrt{\frac{3}{2}}\, e_5.
\end{equation}

\noindent
We see that for this symmetry breaking pattern the standard unification conditions (\ref{Standard_Coupling_Unification}) are not valid. In particular, the value of the Weinberg angle predicted in this case is

\begin{equation}
\sin^2\theta_W = \frac{3}{5},
\end{equation}

\noindent
and the coupling constants are given by the equations

\begin{equation}
\mbox{B-1-1-2:}\qquad \alpha_3 = \alpha_2 = \alpha_5;\qquad \alpha_1^{(Y)} = \frac{3}{2} \alpha_5.
\end{equation}

Calculating the values of the hypercharge (\ref{SU3_SU2_B112_Little_Group_Charge}) for all terms in Eq. (\ref{E8_SU3_SU2_B11_Preliminary}) we obtain the branching rule for the $E_8$ representation $248$,

\begin{eqnarray}
&&\hspace*{-4mm} 248\Big|_{E_8} = 19\times\bm{[1,1](0)} + 8\times[1,1](1/2) +8\times[1,1](-1/2) + 12\times [1,2](0)+ 4\times\bm{[1,2](1/2)}\nonumber\\
&&\hspace*{-4mm}\qquad\quad\ + 4\times\bm{[1,2](-1/2)} + [1,3](0) +8\times [3,1](1/6) + 8\times[\,\xbar{3}\,,1](-1/6) + 6\times \bm{[3,1](-1/3)} \vphantom{\Big(}\nonumber\\
&&\hspace*{-4mm}\qquad\quad\ + 6\times [\,\xbar{3}\,,1](1/3) + \bm{[3,1](2/3)} + [\,\xbar{3}\,,1](-2/3) + 2\times [3,2](-1/3) + 2\times [\,\xbar{3}\,,2](1/3) \vphantom{\Big(}\nonumber\\
&&\hspace*{-4mm}\qquad\quad\ +4\times [3,2](1/6) + 4\times \bm{[\,\xbar{3}\,,2](-1/6)} + [8,1](0)\Big|_{SU_3\times SU_2\times U_1}.
\end{eqnarray}

\noindent
Again, the bold font indicates the representations and hypercharges needed for the MSSM superfields. Certainly, these values of the hypercharge were obtained due to the proper choice of the normalization factor in Eq. (\ref{SU3_SU2_B112_Little_Group_Charge}). However, in this case there are no terms $[1,1](-1)$ corresponding to the right charged leptons, so that this option should be excluded. The most similar representation $[1,1](-1/2)$ has a different value of hypercharge.

The details of the symmetry breaking for the option B-1-1-2 considered here are summarized in Table \ref{Table_Summary_B-1-1-2}.

\bigskip

{\bf B-1-2-1. Vacuum expectation values of $27(-1)\Big|_{E_6\times U_1}$, $16(3)\Big|_{SO_{10}\times U_1}$, and $10(-2)\Big|_{SU_5\times U_1}$.}

\medskip

Now let us consider the option B-1-2. According to Eq. (\ref{248_To_SU5_B12}), the symmetry breaking $SU_5\times U_1\to SU_3\times SU_2\times U_1$ can be realized by vacuum expectation values either of the representations $10(-2)$ (and/or $\,\xbar{10}\,(2)$) or of the representations $10(3)$ (and/or $\,\xbar{10}\,(-3)$). Here we describe the details of the symmetry breaking for the former option B-1-2-1.

The branching rule for the representation $10(-2)$ with respect to the $SU_3\times SU_2\times U_1 \times U_1 $ subgroup (\ref{SU3_SU2_U1^2_Subgroup}) is written as

\begin{equation}\label{10_B121_Branching_Rule}
10(-2)\Big|_{SU_5\times U_1} = \bm{[1,1](-2,6)} + [\,\xbar{3},1](-2,-4) + [3,2](-2,1)\Big|_{SU_3\times SU_2\times U_1\times U_1}.
\end{equation}

From this equation we see that the charge corresponding to the $U_1$ component of the little group $SU_3\times SU_2 \times U_1$ can be chosen as

\begin{equation}\label{SU3_SU2_B121_Little_Group_Charge}
Q_1^{(Y)} = -\frac{1}{30} \Big(3 Q_1^{(5)} + Q_2^{(5)} \Big),
\end{equation}

\noindent
where the normalization factor provides the usual values of the hypercharge for various low energy (super)fields. Taking into account that for the symmetry breaking pattern B-1-2 the coupling constant for the $U_1$ component of $SU_5\times U_1$ is

\begin{equation}
e_1^{(5)} = \frac{e_{5}}{2\sqrt{15}}
\end{equation}

\noindent
we see that now the charge $Q_1^{(5)}$ is an eigenvalue of the operator $2\sqrt{15}\, t_1^{(5)}$. As in the case B-1-1, the charge $Q_2^{(5)}$ can be considered as an eigenvalue of the operator (\ref{Q_2^5_Operator}). Therefore, the charge $Q_1^{(Y)}$ defined by Eq. (\ref{SU3_SU2_B121_Little_Group_Charge}) is an eigenvalue of the operator

\begin{equation}
-\frac{1}{30} \bigg(3 \cdot 2\sqrt{15}\, t_1^{(5)} + 2\sqrt{15}\, t\Big|_{U_1\subset SU_5} \bigg) = \sqrt{\frac{2}{3}} \bigg(-\frac{3}{\sqrt{10}}\, t_1^{(5)} - \frac{1}{\sqrt{10}}\, t\Big|_{U_1\subset SU_5}\bigg).
\end{equation}

\noindent
As earlier, in the right hand side we extracted the operator normalized in the same way as the $SU_5$ generators. Therefore, the coupling constant for the $U_1$ component of the little group $SU_3\times SU_2\times U_1$ is related to the $SU_5$ coupling constant by the equation

\begin{equation}
e_1^{(Y)} = \sqrt{\frac{3}{2}}\, e_5.
\end{equation}

\noindent
Therefore, in this case the gauge coupling constants of the low energy $SU_3\times SU_2\times U_1$ theory take the form

\begin{equation}
\mbox{B-1-2-1:}\qquad \alpha_3=\alpha_2 = \alpha_5;\qquad \alpha_1^{(Y)} = \frac{3}{2} \alpha_5
\end{equation}

\noindent
and coincide with the ones in the option B-1-1-2. In particular, this implies that $\sin^2\theta_W =3/5$ and the standard unification condition (\ref{Standard_Coupling_Unification}) is not satisfied.

\begin{table}[h]
\begin{center}
\begin{tabular}{|c|c|c|c|c|c|c|c|}
\hline
Group $\vphantom{\Big(}$ & \multicolumn{6}{|c|}{Quantum numbers} & Couplings \\
\hline
$E_8$ $\vphantom{\Big(}$ & \multicolumn{6}{|c|}{$\bm{248}$} & $e_8$ \\
\hline
$E_7\times U_1$ $\vphantom{\Big(}$ & $1(0)$ & $1(2)$ & $1(-2)$ & $133(0)$ & $\bm{56(1)}$ & $56(-1)$ & $e_7=e_8/\sqrt{5}$ \\
$\vphantom{\Big(}$                 &        &        &         &          &                &            & $\big(e_1^{(7)} = e_7/4\sqrt{3}\big)$ \\
\hline
$E_6\times U_1$ $\vphantom{\Big(}$ & $1(0)$ & $1(-3)$ & $1(3)$ & $78(0)$          & $\hspace*{0.5mm}\bm{27(-1)}\hspace*{0.5mm}$        & $\,\xbar{27}\,(1)$ & $e_6=e_7/\sqrt{2}$            \\
$\vphantom{\Big(}$                 &        &        &         & $1(0)$           & $\,\xbar{27}\,(-2)$  & $27(2)$  & $\big(e_1^{(6)}=e_6/6\sqrt{2}\big)$ \\
$\vphantom{\Big(}$                 &        &        &         & $\hspace*{0.5mm}\bm{27(-1)}\hspace*{0.5mm}$  & $1(0)$           & $1(0)$          &                               \\
$\vphantom{\Big(}$                 &        &        &         & $\,\xbar{27}\,(1)$  & $1(-3)$           & $1(3)$           &                               \\
\hline
$SO_{10}\times U_1$ $\vphantom{\Big(}$ & \multicolumn{6}{|l|}{$9\times 1(0) + 3\times 1(4) + 3\times 1(-4) + 45(0) +3\times 10(-2)$}
& $e_{10} = e_6/\sqrt{3}$\\
$\vphantom{\Big(}$ & \multicolumn{6}{|l|}{$ + 3\times 10(2) + 3\times 16(-1) + 3\times \,\xbar{16}\,(1) + \bm{16(3)} + \,\xbar{16}\,(-3) $} & $\big(e_1^{(10)} = e_{10}/4\sqrt{3}\big)$\\
\hline
$SU_{5}\times U_1$ $\vphantom{\Big(}$ & \multicolumn{6}{|l|}{$12\times 1(0) + 6\times 1(5) + 6\times 1(-5) + 24(0) + 6\times 5(-1) $} & $e_{5} = e_{10}/\sqrt{2}$\\
$\vphantom{\Big(}$ & \multicolumn{6}{|l|}{$ + 6\times \,\xbar{5}\,(1) + 3\times 5(4) + 3\times \,\xbar{5}\,(-4) +5(-6) + \xbar 5(6)$} & $\big(e_1^{(5)}=e_{5}/2\sqrt{15}\big)$\\
$\vphantom{\Big(}$ & \multicolumn{6}{|l|}{$ + 3\times\bm{{10}(-2)} + 3\times \,\xbar{10}\,(2) + 2\times 10(3) + 2\times \,\xbar{10}\,(-3) $} & \\
\hline
\hspace*{-1mm}$SU_3\times SU_2\times U_1$\hspace*{-1mm} & \multicolumn{6}{|l|}{$19\times\bm{[1,1](0)} + 8\times[1,1](1/2) +8\times[1,1](-1/2)$} & $e_3 = e_2 =e_5\vphantom{\Big(}$ \\
$\vphantom{\Big(}$ & \multicolumn{6}{|l|}{$ + 12\times [1,2](0)+ 4\times\bm{[1,2](1/2)} + 4\times\bm{[1,2](-1/2)} $} & \hspace*{-1mm}$\big(e_1^{(Y)} = e_{5}\sqrt{3/2}\big)$\hspace*{-1mm}\\
$\vphantom{\Big(}$ & \multicolumn{6}{|l|}{$ + [1,3](0) +8\times [3,1](1/6) + 8\times[\,\xbar{3}\,,1](-1/6) $} & $\sin^2\theta_W=3/5$ \\
$\vphantom{\Big(}$ & \multicolumn{6}{|l|}{$ + 6\times \bm{[3,1](-1/3)} + 6\times [\,\xbar{3}\,,1](1/3) + \bm{[3,1](2/3)}  $} & \\
$\vphantom{\Big(}$ & \multicolumn{6}{|l|}{$ + [\,\xbar{3}\,,1](-2/3) + 2\times [3,2](-1/3) + 2\times [\,\xbar{3}\,,2](1/3)  $} & \\
$\vphantom{\Big(}$ & \multicolumn{6}{|l|}{$ +4\times [3,2](1/6) + 4\times \bm{[\,\xbar{3}\,,2](-1/6)} + [8,1](0)$} & \\
\hline
\end{tabular}
\end{center}
\caption{Breaking of 248 into irreducible representations of various subgroups and the corresponding coupling constants for the option {\bf{B-1-2-1}}. In this case the unification condition for the gauge couplings of the low energy $SU_3\times SU_2\times U_1$ theory differs from Eq. (\ref{Standard_Coupling_Unification}), and the representation $[1,1](-1)$ is absent in the final branching rule. The resulting values of couplings and the final branching rule for the representation $248$ coincide with the ones for the option B-1-1-2.}\label{Table_Summary_B-1-2-1}
\end{table}

To obtain the branching rule for the $E_8$ representation $248$ with respect to the little group, we start with the decomposition of this representation with respect to $SU_3\times SU_2\times U_1\times U_1$ subgroup (\ref{SU3_SU2_U1^2_Subgroup}), which can be constructed with the help of Eqs. (\ref{248_To_SU5_B12}) and (\ref{SU5_Branching_Rules}),

\begin{eqnarray}\label{E8_SU3_SU2_B12_Preliminary}
&&\hspace*{-7mm} 248\Big|_{E_8} = 13\times [1,1](0,0) + 2\times [1,1](3,6) + 2\times [1,1](-3,-6) + 3\times [1,1](-2,6)\nonumber\\
&&\hspace*{-7mm}\qquad\quad\ + 3\times [1,1](2,-6) + 6\times [1,1](5,0) + 6\times [1,1](-5,0) + [1,2](-6,3) +[1,2](6,-3) \vphantom{\Big(}\nonumber\\
&&\hspace*{-7mm}\qquad\quad\ + 3\times [1,2](4,3) + 3\times [1,2](-4,-3) + 6\times [1,2](-1,3) + 6\times [1,2](1,-3) \vphantom{\Big(}\nonumber\\
&&\hspace*{-7mm}\qquad\quad\ + [1,3](0,0) + 3\times [3,1](2,4) + 3\times [\,\xbar{3}\,,1](-2,-4) + [3,1](-6,-2) + [\,\xbar{3}\,,1](6,2) \vphantom{\Big(}\nonumber\\
&&\hspace*{-7mm}\qquad\quad\ + 2\times [3,1](-3,4) + 2\times [\,\xbar{3}\,,1](3,-4) + 3\times [3,1](4,-2) + 3\times [\,\xbar{3}\,,1](-4,2)  \vphantom{\Big(}\nonumber\\
&&\hspace*{-7mm}\qquad\quad\ + 6\times [3,1](-1,-2) + 6\times [\,\xbar{3}\,,1](1,2) + [3,2](0,-5) + [\,\xbar{3}\,,2](0,5) + 3\times [3,2](-2,1) \vphantom{\Big(}\nonumber\\
&&\hspace*{-7mm}\qquad\quad\ + 3\times [\,\xbar{3}\,,2](2,-1) + 2\times [3,2](3,1) + 2\times [\,\xbar{3}\,,2](-3,-1)  + [8,1](0,0)\Big|_{SU_3\times SU_2\times U_1\times U_1}.\nonumber\\
\end{eqnarray}

\noindent
Next, for all terms in this expression it is necessary to calculate the charges with respect to the $U_1$ component of the little group $SU_3\times SU_2\times U_1$ using Eq. (\ref{SU3_SU2_B121_Little_Group_Charge}). Then we obtain the decomposition

\begin{eqnarray}
&&\hspace*{-3mm} 248\Big|_{E_8} = 19\times\bm{[1,1](0)} + 8\times[1,1](1/2) +8\times[1,1](-1/2) + 12\times [1,2](0)+ 4\times\bm{[1,2](1/2)}\nonumber\\
&&\hspace*{-3mm}\qquad\quad\ + 4\times\bm{[1,2](-1/2)} + [1,3](0) +8\times [3,1](1/6) + 8\times[\,\xbar{3}\,,1](-1/6) + 6\times \bm{[3,1](-1/3)}\vphantom{\Big(}\nonumber\\
&&\hspace*{-3mm}\qquad\quad\ + 6\times [\,\xbar{3}\,,1](1/3) + \bm{[3,1](2/3)}
+ [\,\xbar{3}\,,1](-2/3) + 2\times [3,2](-1/3) + 2\times [\,\xbar{3}\,,2](1/3) \vphantom{\Big(}\nonumber\\
&&\hspace*{-3mm}\qquad\quad\ +4\times [3,2](1/6) + 4\times \bm{[\,\xbar{3}\,,2](-1/6)} + [8,1](0)\Big|_{SU_3\times SU_2\times U_1},
\end{eqnarray}

\noindent
which exactly coincides with the one for the option B-1-1-2. Therefore, it does not contain the representation $[1,1](-1)$ needed for the right charged leptons and is not suitable for constructing a phenomenologically acceptable model.

Note that, although the gauge coupling unification and the final branching rule for $248$ are the same for the options B-1-1-2 and B-1-2-1, the corresponding symmetry breaking chains are different. However, in this paper we do not discuss the particle content of the resulting theory which essentially depend on such details of symmetry breaking that are not considered here. For instance, it is necessary to present the Lagrangian of the theory and specify all scalar fields which acquire vacuum expectation values. Therefore, strictly speaking, the options B-1-1-2 and B-1-2-1 are not equivalent in spite of the coincidence of the equations for the low energy theory. For the case B-1-2-1 the details of symmetry breaking (discussed in this paper) are collected in Table \ref{Table_Summary_B-1-2-1}.

\bigskip

{\bf B-1-2-2. Vacuum expectation values of $27(-1)\Big|_{E_6\times U_1}$, $16(3)\Big|_{SO_{10}\times U_1}$, and $10(3)\Big|_{SU_5\times U_1}$.}

\medskip

\begin{table}[h]
\begin{center}
\begin{tabular}{|c|c|c|c|c|c|c|c|}
\hline
Group $\vphantom{\Big(}$ & \multicolumn{6}{|c|}{Quantum numbers} & Couplings \\
\hline
$E_8$ $\vphantom{\Big(}$ & \multicolumn{6}{|c|}{$\bm{248}$} & $e_8$ \\
\hline
$E_7\times U_1$ $\vphantom{\Big(}$ & $1(0)$ & $1(2)$ & $1(-2)$ & $133(0)$ & $\bm{56(1)}$ & $56(-1)$ & $e_7=e_8/\sqrt{5}$ \\
$\vphantom{\Big(}$                 &        &        &         &          &                &            & $\big(e_1^{(7)} = e_7/4\sqrt{3}\big)$ \\
\hline
$E_6\times U_1$ $\vphantom{\Big(}$ & $1(0)$ & $1(-3)$ & $1(3)$ & $78(0)$          & $\hspace*{0.5mm}\bm{27(-1)}\hspace*{0.5mm}$        & $\,\xbar{27}\,(1)$ & $e_6=e_7/\sqrt{2}$            \\
$\vphantom{\Big(}$                 &        &        &         & $1(0)$           & $\,\xbar{27}\,(-2)$  & $27(2)$  & $\big(e_1^{(6)}=e_6/6\sqrt{2}\big)$ \\
$\vphantom{\Big(}$                 &        &        &         & $\hspace*{0.5mm}\bm{27(-1)}\hspace*{0.5mm}$  & $1(0)$           & $1(0)$          &                               \\
$\vphantom{\Big(}$                 &        &        &         & $\,\xbar{27}\,(1)$  & $1(-3)$           & $1(3)$           &                               \\
\hline
$SO_{10}\times U_1$ $\vphantom{\Big(}$ & \multicolumn{6}{|l|}{$9\times 1(0) + 3\times 1(4) + 3\times 1(-4) + 45(0) +3\times 10(-2)$}
& $e_{10} = e_6/\sqrt{3}$\\
$\vphantom{\Big(}$ & \multicolumn{6}{|l|}{$ + 3\times 10(2) + 3\times 16(-1) + 3\times \,\xbar{16}\,(1) + \bm{16(3)} + \,\xbar{16}\,(-3) $} & $\big(e_1^{(10)} = e_{10}/4\sqrt{3}\big)$\\
\hline
$SU_{5}\times U_1$ $\vphantom{\Big(}$ & \multicolumn{6}{|l|}{$12\times 1(0) + 6\times 1(5) + 6\times 1(-5) + 24(0) + 6\times 5(-1) $} & $e_{5} = e_{10}/\sqrt{2}$\\
$\vphantom{\Big(}$ & \multicolumn{6}{|l|}{$ + 6\times \,\xbar{5}\,(1) + 3\times 5(4) + 3\times \,\xbar{5}\,(-4) +5(-6) + \,\xbar{5}\,(6)$} & $\big(e_1^{(5)}=e_{5}/2\sqrt{15}\big)$\\
$\vphantom{\Big(}$ & \multicolumn{6}{|l|}{$ + 3\times {10}(-2) + 3\times \,\xbar{10}\,(2) + 2\times \bm{10(3)} + 2\times \,\xbar{10}\,(-3) $} & \\
\hline
\hspace*{-1mm}$SU_3\times SU_2\times U_1$\hspace*{-1mm} & \multicolumn{6}{|l|}{$17\times\bm{[1,1](0)} + 9\times[1,1](1/3) +9\times[1,1](-1/3) $} & $e_3 = e_2 =e_5\vphantom{\Big(}$ \\
$\vphantom{\Big(}$ & \multicolumn{6}{|l|}{$ + \bm{[1,2](1/2)} + \bm{[1,2](-1/2)} + 9\times [1,2](1/6) $} & \hspace*{-1mm}$\big(e_1^{(Y)} = e_{5}\sqrt{3}\big)$\hspace*{-1mm}\\
$\vphantom{\Big(}$ & \multicolumn{6}{|l|}{$ + 9\times [1,2](-1/6) + [1,3](0) + 9\times [3,1](0) + 9\times[\,\xbar{3}\,,1](0)$} & $\sin^2\theta_W=3/4$ \\
$\vphantom{\Big(}$ & \multicolumn{6}{|l|}{$ +3\times [3,1](1/3) + 3\times[\,\xbar{3}\,,1](-1/3) +3\times \bm{[3,1](-1/3)}  $} & \\
$\vphantom{\Big(}$ & \multicolumn{6}{|l|}{$ + 3\times[\,\xbar{3}\,,1](1/3) + 3\times [3,2](-1/6) + 3\times [\,\xbar{3}\,,2](1/6) $} & \\
$\vphantom{\Big(}$ & \multicolumn{6}{|l|}{$ + 3\times [3,2](1/6) + 3\times \bm{[\,\xbar{3}\,,2](-1/6)} + [8,1](0)$} & \\
\hline
\end{tabular}
\end{center}
\caption{Breaking of 248 into irreducible representations of various subgroups and the corresponding coupling constants for the option {\bf{B-1-2-2}}. In this case the unification condition for the gauge couplings of the low energy $SU_3\times SU_2\times U_1$ theory differs from Eq. (\ref{Standard_Coupling_Unification}), and the representations $[1,1](-1)$ and $[3,1](2/3)$ are absent in the final branching rule.}\label{Table_Summary_B-1-2-2}
\end{table}

For this case the $SU_5\times U_1$ symmetry is broken down to $SU_3\times SU_2\times U_1$ by vacuum expectation values of the representations $10(3)$ (and/or $\,\xbar{10}\,(-3)$) present in the decomposion (\ref{248_To_SU5_B12}). According to Eq. (\ref{SU5_Branching_Rules}), the branching rule for the representation $10(3)$ with respect to the subgroup (\ref{SU3_SU2_U1^2_Subgroup}) reads as

\begin{equation}\label{10_B122_Branching_Rule}
10(3)\Big|_{SU_5\times U_1} = \bm{[1,1](3,6)} + [\,\xbar{3},1](3,-4) + [3,2](3,1)\Big|_{SU_3\times SU_2\times U_1\times U_1}.
\end{equation}

\noindent
This implies that the charge corresponding to the $U_1$ component of the little group $SU_3\times SU_2\times U_1$ can be chosen in the form

\begin{equation}\label{SU3_SU2_B122_Little_Group_Charge}
Q_1^{(Y)} = \frac{1}{30} \Big(2 Q_1^{(5)} - Q_2^{(5)} \Big),
\end{equation}

\noindent
where (as in the previous cases) the factor $1/30$ is included in order that the resulting decomposition of $248$ contains the standard values of hypercharge for various MSSM superfields. Similarly to the case B-1-2-1, the little group charge $Q_1^{(Y)}$ given by Eq. (\ref{SU3_SU2_B122_Little_Group_Charge}) is obtained as an eigenvalue of the operator

\begin{equation}
\frac{1}{30} \bigg(2 \cdot 2\sqrt{15}\, t_1^{(5)} - 2\sqrt{15}\, t\Big|_{U_1\subset SU_5} \bigg) =\frac{1}{\sqrt{3}} \bigg(\frac{2}{\sqrt{5}}\, t_1^{(5)} - \frac{1}{\sqrt{5}}\, t\Big|_{U_1\subset SU_5}\bigg),
\end{equation}

\noindent
where the rightmost operator in the round brackets is normalized in the same way as the generators of the group $SU_5$. Therefore, the gauge coupling corresponding to the $U_1$ component of the little group is given by the expression

\begin{equation}
e_1^{(Y)} = \sqrt{3}\, e_5,
\end{equation}

\begin{table}[h]
\begin{center}
\begin{tabular}{|c|c|c|c|c|c|c|c|}
\hline
Group $\vphantom{\Big(}$ & \multicolumn{6}{|c|}{Quantum numbers} & Couplings \\
\hline
$E_8$ $\vphantom{\Big(}$ & \multicolumn{6}{|c|}{$\bm{248}$} & $e_8$ \\
\hline
$E_7\times U_1$ $\vphantom{\Big(}$ & $1(0)$ & $1(2)$ & $1(-2)$ & $133(0)$ & $\bm{56(1)}$ & $56(-1)$ & $e_7=e_8/\sqrt{5}$ \\
$\vphantom{\Big(}$                 &        &        &         &          &                &            & $\big(e_1^{(7)} = e_7/4\sqrt{3}\big)$ \\
\hline
$E_6\times U_1$ $\vphantom{\Big(}$ & $1(0)$ & $1(-3)$ & $1(3)$ & $78(0)$                                & $\hspace*{0.5mm}27(-1)\hspace*{0.5mm}$    & $\,\xbar{27}\,(1)$                       & $e_6=e_7/\sqrt{2}$             \\
$\vphantom{\Big(}$                 &        &        &         & $1(0)$                                 & $\,\xbar{27}\,(-2)$  & $\hspace*{0.5mm}\bm{27(2)}\hspace*{0.5mm}$  & $\big(e_1^{(6)}=e_6/6\sqrt{2}\big)$ \\
$\vphantom{\Big(}$                 &        &        &         & $\hspace*{0.5mm}27(-1)\hspace*{0.5mm}$   & $1(0)$          & $1(0)$                               &                                \\
$\vphantom{\Big(}$                 &        &        &         & $\,\xbar{27}\,(1)$                         & $1(-3)$          & $1(3)$                                &                                \\
\hline
$SO_{10}\times U_1$ $\vphantom{\Big(}$ & \multicolumn{6}{|l|}{$7\times 1(0) + 4\times 1(2) + 4\times 1(-2) +4\times 10(0) + 10(2)$}
& $e_{10} = e_6/\sqrt{3}$\\
$\vphantom{\Big(}$ & \multicolumn{6}{|l|}{$ + 10(-2) + 2\times \bm{16(1)} + 2\times \,\xbar{16}\,(-1) + 2\times 16(-1) $} & $\big(e_1^{(10)} = e_{10}/4\big)$\\
$\vphantom{\Big(}$ & \multicolumn{6}{|l|}{$ + 2\times \,\xbar{16}\,(1) + 45(0) $} &  \\
\hline
$SU_{5}\times U_1$ $\vphantom{\Big(}$ & \multicolumn{6}{|l|}{$12\times 1(0) + 6\times 1(5) + 6\times 1(-5) + 24(0) + 6\times 5(-1) $} & $e_{5} = e_{10}/\sqrt{2}$\\
$\vphantom{\Big(}$ & \multicolumn{6}{|l|}{$ + 6\times \,\xbar{5}\,(1) + 3\times 5(4) + 3\times \,\xbar{5}\,(-4) + 5(-6) + \,\xbar{5}\,(6) $} & $\big(e_1^{(5)}=e_{5}/2\sqrt{15}\big)$\\
$\vphantom{\Big(}$ & \multicolumn{6}{|l|}{$ + 3\times \bm{10(-2)} + 3\times \,\xbar{10}\,(2) + 2\times 10(3) + 2\times \,\xbar{10}\,(-3)  $} & \\
\hline
\hspace*{-1mm}$SU_3\times SU_2\times U_1$\hspace*{-1mm} & \multicolumn{6}{|l|}{$19\times\bm{[1,1](0)} + 8\times[1,1](1/2) + 8\times[1,1](-1/2) $} & $e_3 = e_2 =e_5\vphantom{\Big(}$ \\
$\vphantom{\Big(}$ & \multicolumn{6}{|l|}{$ + 12\times[1,2](0) + 4\times\bm{[1,2](1/2)} + 4\times\bm{[1,2](-1/2)} $} & \hspace*{-1mm}$\big(e_1^{(Y)} = e_{5} \sqrt{3/2}\big)$\hspace*{-1mm}\\
$\vphantom{\Big(}$ & \multicolumn{6}{|l|}{$ + [1,3](0) +8\times [3,1](1/6) + 8\times [\,\xbar{3}\,,1](-1/6) $} & $\sin^2\theta_W=3/5$ \\
$\vphantom{\Big(}$ & \multicolumn{6}{|l|}{$  + 6\times\bm{[3,1](-1/3)} + 6\times[\,\xbar{3}\,,1](1/3) + \bm{[3,1](2/3)} $} & \\
$\vphantom{\Big(}$ & \multicolumn{6}{|l|}{$ + [\,\xbar{3}\,,1](-2/3) + 2\times [3,2](-1/3) +2\times [\,\xbar{3}\,,2](1/3)  $} & \\
$\vphantom{\Big(}$ & \multicolumn{6}{|l|}{$ +4\times [3,2](1/6) +4\times \bm{[\,\xbar{3}\,,2](-1/6)} + [8,1](0)$} & \\
\hline
\end{tabular}
\end{center}
\caption{Breaking of 248 into irreducible representations of various subgroups appearing in the symmetry breaking pattern {\bf{B-2-1-1}}. The final branching rule and the final values of the coupling constants are the same as in the options B-1-1-2 and B-1-2-1.}\label{Table_Summary_B-2-1-1}
\end{table}

\noindent
so that in the case under consideration the gauge couplings of the low energy theory can be expressed in terms of the $SU_5$ coupling constant as

\begin{equation}
\mbox{B-1-2-2:}\qquad \alpha_3=\alpha_2 = \alpha_5;\qquad \alpha_1^{(Y)} = 3 \alpha_5.
\end{equation}

\noindent
This implies that the value of the Weinberg angle for the symmetry breaking pattern B-1-2-2 is

\begin{equation}
\sin^2\theta_W = \frac{3}{4}.
\end{equation}

\begin{table}[h]
\begin{center}
\begin{tabular}{|c|c|c|c|c|c|c|c|}
\hline
Group $\vphantom{\Big(}$ & \multicolumn{6}{|c|}{Quantum numbers} & Couplings \\
\hline
$E_8$ $\vphantom{\Big(}$ & \multicolumn{6}{|c|}{$\bm{248}$} & $e_8$ \\
\hline
$E_7\times U_1$ $\vphantom{\Big(}$ & $1(0)$ & $1(2)$ & $1(-2)$ & $133(0)$ & $\bm{56(1)}$ & $56(-1)$ & $e_7=e_8/\sqrt{5}$ \\
$\vphantom{\Big(}$                 &        &        &         &          &                &            & $\big(e_1^{(7)} = e_7/4\sqrt{3}\big)$ \\
\hline
$E_6\times U_1$ $\vphantom{\Big(}$ & $1(0)$ & $1(-3)$ & $1(3)$ & $78(0)$          & $\hspace*{2mm}27(-1)\hspace*{2mm}$        & $\,\xbar{27}\,(1)$ & $e_6=e_7/\sqrt{2}$            \\
$\vphantom{\Big(}$                 &        &        &         & $1(0)$           & $\,\xbar{27}\,(-2)$  & $\hspace*{0.5mm}\bm{27(2)}\hspace*{0.5mm}$  & $\big(e_1^{(6)}=e_6/6\sqrt{2}\big)$ \\
$\vphantom{\Big(}$                 &        &        &         & $\hspace*{2mm}27(-1)\hspace*{2mm}$  & $1(0)$           & $1(0)$          &                               \\
$\vphantom{\Big(}$                 &        &        &         & $\,\xbar{27}\,(1)$  & $1(-3)$           & $1(3)$           &                               \\
\hline
$SO_{10}\times U_1$ $\vphantom{\Big(}$ & \multicolumn{6}{|l|}{$7\times 1(0) + 4\times 1(2) + 4\times 1(-2) +4\times 10(0) + 10(2)$}
& $e_{10} = e_6/\sqrt{3}$\\
$\vphantom{\Big(}$ & \multicolumn{6}{|l|}{$ + 10(-2) + 2\times \bm{16(1)} + 2\times \,\xbar{16}\,(-1) + 2\times 16(-1) $} & $\big(e_1^{(10)} = e_{10}/4\big)$\\
$\vphantom{\Big(}$ & \multicolumn{6}{|l|}{$ + 2\times \,\xbar{16}\,(1) + 45(0) $} &  \\
\hline
$SU_{5}\times U_1$ $\vphantom{\Big(}$ & \multicolumn{6}{|l|}{$12\times 1(0) + 6\times 1(5) + 6\times 1(-5) + 24(0) + 6\times 5(-1) $} & $e_{5} = e_{10}/\sqrt{2}$\\
$\vphantom{\Big(}$ & \multicolumn{6}{|l|}{$ + 6\times \,\xbar{5}\,(1) + 3\times 5(4) + 3\times \,\xbar{5}\,(-4) + 5(-6) + \,\xbar{5}\,(6) $} & $\big(e_1^{(5)}=e_{5}/2\sqrt{15}\big)$\\
$\vphantom{\Big(}$ & \multicolumn{6}{|l|}{$ + 3\times 10(-2) + 3\times \,\xbar{10}\,(2) + 2\times \bm{10(3)} + 2\times \,\xbar{10}\,(-3) $} & \\
\hline
\hspace*{-1mm}$SU_3\times SU_2\times U_1$\hspace*{-1mm} & \multicolumn{6}{|l|}{$17\times\bm{[1,1](0)} + 9\times[1,1](1/3) +9\times[1,1](-1/3) $} & $e_3 = e_2 =e_5\vphantom{\Big(}$ \\
$\vphantom{\Big(}$ & \multicolumn{6}{|l|}{$ + [1,3](0) + 9\times[1,2](-1/6) + 9\times[1,2](1/6) $} & \hspace*{-1mm}$\big(e_1^{(Y)} = e_{5} \sqrt{3}\big)$\hspace*{-1mm}\\
$\vphantom{\Big(}$ & \multicolumn{6}{|l|}{$ + \bm{[1,2](-1/2)} + \bm{[1,2](1/2)} +9\times [3,1](0) + 9\times [\,\xbar{3}\,,1](0) $} & $\sin^2\theta_W=3/4$ \\
$\vphantom{\Big(}$ & \multicolumn{6}{|l|}{$  + 3\times[3,1](1/3) + 3\times[\,\xbar{3}\,,1](-1/3) + 3\times\bm{[3,1](-1/3)} $} & \\
$\vphantom{\Big(}$ & \multicolumn{6}{|l|}{$ + 3\times[\,\xbar{3}\,,1](1/3) + 3\times [3,2](-1/6) +3\times [\,\xbar{3}\,,2](1/6)  $} & \\
$\vphantom{\Big(}$ & \multicolumn{6}{|l|}{$ +3\times [3,2](1/6) +3\times \bm{[\,\xbar{3}\,,2](-1/6)} + [8,1](0)$} & \\
\hline
\end{tabular}
\end{center}
\caption{Breaking of 248 into irreducible representations of various subgroups appearing in the symmetry breaking pattern {\bf{B-2-1-2}}. The final branching rule and the final values of the coupling constants are the same as in the option B-1-2-2.}\label{Table_Summary_B-2-1-2}
\end{table}

Calculating the charges for the $U_1$ component of little group with the help of Eq. (\ref{SU3_SU2_B122_Little_Group_Charge}) for all terms in Eq. (\ref{E8_SU3_SU2_B12_Preliminary}) we obtain the decomposition of the $E_8$ representation $248$ with respect to the little group $SU_3\times SU_2\times U_1$,

\begin{eqnarray}
&& 248\Big|_{E_8} = 17\times\bm{[1,1](0)} + 9\times[1,1](1/3) + 9\times[1,1](-1/3) + \bm{[1,2](1/2)} + \bm{[1,2](-1/2)} \nonumber\\
&&\qquad\quad\ + 9\times [1,2](1/6) + 9\times [1,2](-1/6) + [1,3](0) + 9\times [3,1](0) + 9\times[\,\xbar 3\,,1](0)\vphantom{\Big(}\nonumber\\
&&\qquad\quad\ + 3\times [3,1](1/3) + 3\times[\,\xbar{3}\,,1](-1/3) +3\times \bm{[3,1](-1/3)} + 3\times[\,\xbar{3}\,,1](1/3) \vphantom{\Big(}\nonumber\\
&&\qquad\quad\ + 3\times [3,2](-1/6) + 3\times [\,\xbar{3}\,,2](1/6) + 3\times [3,2](1/6) + 3\times \bm{[\,\xbar{3}\,,2](-1/6)}\vphantom{\Big(}\nonumber\\
&&\qquad\quad\ + [8,1](0)\Big|_{SU_3\times SU_2\times U_1}.
\end{eqnarray}

\noindent
We see that this expression does not contain all parts needed for various MSSM superfields. The factor in Eq. (\ref{SU3_SU2_B122_Little_Group_Charge}) was chosen in such a way that the above expression contains a part $[\,\xbar{3}\,,2](-1/6)$ corresponding to the (charge conjugated) left quarks. However, in this case the representations $[1,1](-1)$ and $[3,1](2/3)$ needed for the right charged leptons and for the right upper quarks, respectively, are absent.  Therefore, this option for the symmetry breaking should also be excluded.

The details of the symmetry breaking for the case B-1-2-2 are collected in Table \ref{Table_Summary_B-1-2-2}.

\bigskip

{\bf B-2-1-1. Vacuum expectation values of $27(2)\Big|_{E_6\times U_1}$, $16(1)\Big|_{SO_{10}\times U_1}$, and $10(-2)\Big|_{SU_5\times U_1}$.}

\medskip

As we have mentioned above, the option B-2-1 leads to the same values of the $SU_5\times U_1$ coupling constants and to the same branching rule for the representation $248$ of $E_8$ with respect to $SU_5\times U_1$ as the option B-1-2. That is why the last symmetry breaking $SU_5\times U_1 \to SU_3\times SU_2 \times U_1$ now occurs exactly in the same way as for the option B-1-2-1. However, the previous symmetry breakings $E_6\times U_1 \to SO_{10}\times U_1 \to SU_5\times U_1$ for the options B-2-1-1 and B-1-2-1 are different. The details of the symmetry breaking chain for the option B-2-1-1 are presented in Table \ref{Table_Summary_B-2-1-1}. We see that in this case the values of gauge couplings for the low energy $SU_3\times SU_2 \times U_1$ theory do not satisfy the standard conditions (\ref{Standard_Coupling_Unification}), and the particle content does not match MSSM.

\bigskip

{\bf B-2-1-2. Vacuum expectation values of $27(2)\Big|_{E_6\times U_1}$, $16(1)\Big|_{SO_{10}\times U_1}$, and $10(3)\Big|_{SU_5\times U_1}$.}

\medskip

Because the values of $SU_5\times U_1$ coupling constants and the branching rule for the representation $248$ of $E_8$ with respect to $SU_5\times U_1$ are the same for the options B-2-1 and B-1-2, in this case the symmetry breaking $SU_5\times U_1 \to SU_3\times SU_2\times U_1$ coincides with the one for the option B-1-2-2. The whole chain of the symmetry breaking including values of the coupling constants and branching rules for the representation $248$ is described in Table \ref{Table_Summary_B-2-1-2}. Again, for this symmetry breaking pattern the unification conditions are different from Eq. (\ref{Standard_Coupling_Unification}), and the particle content does not match MSSM.

\section{Conclusion}
\hspace*{\parindent}

In this paper we investigate a possibility of Grand Unification on the base of a theory with the gauge group $E_8$ assuming that the symmetry breaking is caused by vacuum expectation values of the fields originating from the representation(s) 248 (without involving higher representations). This produces the symmetry breaking chain (\ref{Symmetry_Breaking}),

\begin{equation}
E_8\to E_7\times U_1 \to E_6\times U_1 \to SO_{10}\times U_1 \to SU_5 \times U_1 \to SU_3 \times SU_2 \times U_1,
\end{equation}

\noindent
where the $U_1$ gauge transformations at a certain step are nontrivial combinations of the $U_1$ transformations and the transformations of a simple group at the previous step. In particular, we assume that all $U_1$ groups are different and do not consider the options when a certain $U_1$ group is a subgroup of a simple group at the previous step. (One of such options appears, e.g., if the $SU_5$ symmetry is broken by a vacuum expectation value of the Higgs field in the adjoint representation $24$ down to $SU_3\times SU_2\times U_1$, and the $U_1$ factor in $SU_5\times U_1$ is broken by an $SU_5$ singlet with a nontrivial $U_1$ charge.) All these options are also possible and interesting, and their number is rather large.

Note that although the original $E_8$ theory is vectorlike, it seems possible to obtain a chiral low energy theory with the help of the considered symmetry breaking chain, because vacuum expectation values are acquired by scalar fields with nontrivial $U_1$ charges. This can produce asymmetry between conjugated representations (which always appear symmetrically in the branching rules of the representation $248$), so that the parity breaking can be spontaneous, provided that an appropriate dynamic mechanism is built. However, this issue was not addressed in this paper. Here the main purpose is to investigate if one can achieve the standard unification of gauge couplings corresponding to $\alpha_3=\alpha_2$ and $\sin^2\theta_W=3/8$ at the scale of unification (which certainly occurs at very high energies). In principle, this implies that we deal with supersymmetric theories, because only in the supersymmetric case this unification agrees with the experimental data \cite{Ellis:1990wk,Amaldi:1991cn,Langacker:1991an}. Note that the presence of several symmetry breaking scales certainly change the evolution of running couplings. Nevertheless, if all symmetry breaking scales are close to each other, then the corresponding corrections can possibly be negligible and do not affect the gauge coupling unification. In this paper we did not consider quantum corrections and simply tried to find such a symmetry breaking pattern under which the gauge couplings of the $SU_3\times SU_2\times U_1$ theory at the classical level are related as $e_3=e_2=e_1^{(Y)}\sqrt{5/3}$.

It was revealed that the considered symmetry breaking pattern can be realized by 6 different ways. Only one of them (B-1-1-1) gives the standard unification conditions (\ref{Standard_Coupling_Unification}). Wonderfully, only for this option the decomposition of the representation $248$ contains all parts required to accommodate the MSSM superfields. It is interesting that in this option vacuum expectation values are acquired by scalar fields in the representations 27 of $E_6$, 16 of $SO_{10}$, and $10$ of $SU_5$ with the least possible absolute values of the $U_1$ charge. Probably, this fact and the similarity between all symmetry breakings imply the existence of a certain mechanism for the chain symmetry breaking analogous to the one proposed in \cite{Boos:2021hcb}. Note that for its implementation it is not enough to have only one representation $248$, so that a possibility of using (divergence-free) ${\cal N}=4$ supersymmetric Yang--Mills theory with the gauge group $E_8$ may be considered.

\section*{Acknowledgments}
\hspace*{\parindent}

The author is very grateful to Prof. K.Patel and Prof. R.Maji for indicating some important references.

This work has been supported by Russian Science Foundation, grant No. 21-12-00129.

\appendix

\section*{Appendix}

\section{The branching rules for the fundamental and adjoint representations of $E_7$ with respect to the subgroup $E_6\times U_1$}
\hspace*{\parindent}\label{Appendix_E7_Branching}

The branching rules for the $E_7$ representations with respect to the subgroup $E_6\times U_1$ presented in \cite{Slansky:1981yr} do not contain $U_1$ charges. That is why here we discuss how these charges can be found and calculate them for two lowest $E_7$ representations in which we are interested in.

According to \cite{Slansky:1981yr}, the group $E_7$ contains a subgroup $SO_{12}\times SO_3$. The group $SO_{12}$ in turn contains the subgroup $SO_{10}\times U_1$, so that we obtain the embedding

\begin{equation}\label{E7_U1^2_Gamma_Subgroup}
E_7 \supset SO_{12}\times SO_3 \supset (SO_{10}\times \underbrace{U_1}_{\gamma_2}) \times \underbrace{U_1}_{\gamma_1},
\end{equation}

\noindent
where $\gamma_1$ and $\gamma_2$ are the real parameters of the corresponding $U_1$ transformations. The $E_7$ branching rules for the representations $56$ and $133$ with respect to the subgroup $SO_{12}\times SO_3$ are given by Eq. (\ref{E7_Branching_Rules}). The $SO_{12}$ branching rules with respect to the subgroup $SO_{10}\times U_1$ can also be found in \cite{Slansky:1981yr} and have the form

\begin{eqnarray}\label{SO12_Branching_Rules}
&& 12\Big|_{SO_{12}} = 1(2) + 1(-2) + 10(0)\Big|_{SO_{10}\times U_1};\nonumber\\
&& 32\Big|_{SO_{12}} = 16(1) + \,\xbar{16}\,(-1)\Big|_{SO_{10}\times U_1};\nonumber\\
&& 32'\Big|_{SO_{12}} = 16(-1) + \,\xbar{16}\,(1)\Big|_{SO_{10}\times U_1};\nonumber\\
&& 66\Big|_{SO_{12}} = 1(0) + 10(2) + 10(-2) + 45(0)\Big|_{SO_{10}\times U_1},
\end{eqnarray}

\noindent
where the numbers in brackets are the charges with respect to the $U_1$ group parameterized by $\gamma_2$. The $U_1$ charges for various parts of $SO_3$ representations can also be easily found from the theory of angular momentum. Taking into account that $2J_3$ goes from $-2J$ to $2J$ with the step 2 and using Eqs. (\ref{E7_Branching_Rules}) and (\ref{SO12_Branching_Rules}) we conclude that the fundamental and adjoint representations of $E_7$ are decomposed into irreducible representations of the subgroup (\ref{E7_U1^2_Gamma_Subgroup}) according to the branching rules

\begin{eqnarray}\label{E7_Gamma_Branching_Rules}
&&\hspace*{-7mm} 56\Big|_{E_7} = [12,2] + [32,1]\Big|_{SO_{12}\times SO_3} = 1(1,2) + 1(-1,2) + 1(1,-2) + 1(-1,-2) + 10(1,0) \qquad\nonumber\\
&&\hspace*{-7mm}\qquad\quad + 10(-1,0) + 16(0,1) + \,\xbar{16}\,(0,-1)\Big|_{SO_{10}\times U_1\times U_1};\qquad\vphantom{1_{p_p}}\nonumber\\
&&\hspace*{-7mm} 133\Big|_{E_7} = [1,3] + [66,1] + [32',2]\Big|_{SO_{12}\times SO_3} = 1(2,0)  + 1(0,0) + 1(-2,0) + 1(0,0) + 10(0,2) \vphantom{1^{d^d}} \nonumber\\
&&\hspace*{-7mm}\qquad\quad + 10(0,-2) + 45(0,0) + 16(1,-1)  + \,\xbar{16}\,(1,1) + 16(-1,-1) + \,\xbar{16}\,(-1,1) \Big|_{SO_{10}\times U_1\times U_1}.\nonumber\\
\end{eqnarray}

\noindent
Here the first number in the round brackets is the charge $Q_{\gamma_1}$ corresponding to the $U_1$ subgroup parameterized by $\gamma_1$, and the second number is the charge $Q_{\gamma_2}$ corresponding to the $U_1$ subgroup parameterized by $\gamma_2$.

From the other side, according to \cite{Slansky:1981yr}, the group $E_7$ contains the subgroup $E_6\times U_1$, and the group $E_6$ contains the subgroup $SO_{10}\times U_1$. Therefore, we obtain another embedding

\begin{equation}\label{E7_U1^2_Delta_Subgroup}
E_7 \supset E_6\times \underbrace{U_1}_{\delta_1} \supset (SO_{10}\times \underbrace{U_1}_{\delta_2}) \times \underbrace{U_1}_{\delta_1},
\end{equation}

\noindent
where the parameters of the $U_1$ groups were denoted by $\delta_1$ and $\delta_2$. The branching rules for the $E_7$ representations $56$ and $133$ with respect to the $E_6$ subgroup can be found in \cite{Slansky:1981yr} and are rather evident. However, it is not so trivial to find the corresponding $U_1$ charges. For this purpose we assume that

\begin{eqnarray}\label{E7_To_E6_U1_Branching_Tules}
&& 56\Big|_{E_7} = 27(1) + \,\xbar{27}\,(-1) + 1(x_1) + 1(-x_1)\Big|_{E_6\times U_1};\nonumber\\
&& 133\Big|_{E_7} = 1(x_2) + 27(x_3) + \,\xbar{27}\,(-x_3) + 78(x_4)\Big|_{E_6\times U_1},\qquad
\end{eqnarray}

\noindent
where in the brackets we write the charge corresponding to the $U_1$ subgroup parameterized by $\delta_1$, and the constants $x_i$ are to be calculated. (The $U_1$ charge of the $E_6$ representation $27$ in the first equation can be chosen in arbitrary way. Setting it to 1 we simply specify its normalization.) Using Eq. (\ref{E6_Branching_Rules}) we obtain the branching rules with respect to the subgroup (\ref{E7_U1^2_Delta_Subgroup}) in the form

\begin{eqnarray}\label{E7_Delta_Branching_Rules}
&& 56\Big|_{E_7} = 1(1,4) +10(1,-2) +16(1,1) + 1(-1,-4) + 10(-1,2) + \,\xbar{16}\,(-1,-1) \nonumber\\
&&\qquad\quad\ + 1(x_1,0) + 1(-x_1,0)\Big|_{SO_{10}\times U_1\times U_1};\nonumber\\
&& 133\Big|_{E_7} = 1(x_2,0) + 1(x_3,4) + 10(x_3,-2) + 16(x_3,1) + 1(-x_3,-4) + 10(-x_3,2)\nonumber\\
&&\qquad\quad\ +\,\xbar{16}\,(-x_3,-1) + 1(x_4,0) + 16(x_4,-3) + \,\xbar{16}\,(x_4,3) + 45(x_4,0)\Big|_{SO_{10}\times U_1\times U_1},\qquad
\end{eqnarray}

\noindent
where the numbers in the brackets are the charges $Q_{\delta_1}$ and $Q_{\delta_2}$ with respect to the $U_1$ groups parameterized by $\delta_1$ and $\delta_2$, respectively.

Comparing Eqs. (\ref{E7_Gamma_Branching_Rules}) and (\ref{E7_Delta_Branching_Rules}) we see that (as expected) the $SO_{10}$ representations coincide, but their $U_1$ charges are different. The matter is that the $U_1$ subgroups parameterized by $\gamma_{1,2}$ and $\delta_{1,2}$ are different. However, the $U_1$ charges coincide if we make the identification

\begin{equation}\label{Q_Relations}
\left\{
\begin{array}{l}
Q_{\delta_1} = Q_{\gamma_1} + Q_{\gamma_2};\vphantom{\int\limits_x}\\
Q_{\delta_2} = -2 Q_{\gamma_1} + Q_{\gamma_2}.\vphantom{\int}
\end{array}
\right.
\end{equation}

\noindent
This implies that under the $U_1\times U_1$ transformations a field $\varphi$ changes as

\begin{equation}
\varphi \to \exp\Big(iQ_{\delta_1} \delta_1 + i Q_{\delta_2}\delta_2\Big)\varphi = \exp\Big(i Q_{\gamma_1} (\delta_1 - 2 \delta_2) + iQ_{\gamma_2}(\delta_1 + \delta_2)\Big)\varphi,
\end{equation}

\noindent
so that the parameters of the $U_1\times U_1$ subgroups in (\ref{E7_U1^2_Gamma_Subgroup}) and (\ref{E7_U1^2_Delta_Subgroup}) are related as

\begin{equation}
\left\{
\begin{array}{l}
\gamma_1 = \delta_1 -2\delta_2;\vphantom{\int\limits_x}\\
\gamma_2 = \delta_1 + \delta_2,\vphantom{\int}
\end{array}
\right. \qquad\qquad
\left\{
\begin{array}{l}
\delta_1 = \smash{\displaystyle (\gamma_1 + 2\gamma_2)/3;}\vphantom{\int\limits_p}\\
\delta_2 = \smash{\displaystyle (-\gamma_1 + \gamma_2)/3.}\vphantom{\int}
\end{array}
\right.
\end{equation}

\noindent
Using Eq. (\ref{Q_Relations}) and comparing Eqs. (\ref{E7_Gamma_Branching_Rules}) and (\ref{E7_Delta_Branching_Rules}) we obtain

\begin{equation}
x_1=3;\qquad x_2=0;\qquad x_3=-2;\qquad x_4=0.
\end{equation}

\noindent
Substituting these values into Eq. (\ref{E7_To_E6_U1_Branching_Tules}) we obtain the required branching rules

\begin{eqnarray}
&& 56\Big|_{E_7} = 27(1) + \,\xbar{27}\,(-1) + 1(3) + 1(-3)\Big|_{E_6\times U_1};\\
&& 133\Big|_{E_7} = 1(0) + 27(-2) + \,\xbar{27}\,(2) + 78(0)\Big|_{E_6\times U_1},\qquad
\end{eqnarray}

\noindent
which coincide with Eq. (\ref{E7_Branching_Rules_To_E6}).

\end{document}